\documentclass[eqsecnum,preprint,12pt]{revtex4-1}
\usepackage{graphicx,subfigure,multirow}

\begin{document}

\title{Calculation of renormalized fermion effective actions in radially symmetric non-Abelian backgrounds}
\author{Jin Hur} \email{hurjin@add.re.kr}
\affiliation{Agency for Defense Development, Jochiwongil 462, Yuseong, Daejeon 305-152, Korea}
\author{Choonkyu Lee} \email{cklee@phya.snu.ac.kr}
\affiliation{Department of Physics and Astronomy and Center for Theoretical Physics\\ Seoul National University, Seoul 151-742, Korea}
\author{Hyunsoo Min} \email{hsmin@dirac.uos.ac.kr}
\affiliation{Department of Physics, University of Seoul, Seoul 130-743, Korea}

\begin{abstract}
Our recent method to calculate renormalized functional determinants, the partial wave cutoff method, is extended for the evaluation of 4-D fermion one-loop effective action with arbitrary mass in certain types of radially symmetric, non-Abelian, background gauge fields (including instanton-like and instanton-antiinstanton-like configurations). A detailed study on functional determinants for matrix-valued radial differential operators is presented, explicating both our analytic treatment on the high partial wave contribution and the application of the generalized Gel'fand-Yaglom formula to determine the low partial wave contribution. In general, some numerical work is needed for the low partial wave part. In the massless limit, however, the factorizable nature of our partial-wave radial differential operators can be exploited to evaluate semi-analytically even the low partial wave part, and we thus have the full fermion effective action calculated explicitly in a class of non-Abelian background gauge fields. With nonzero mass, we also perform necessary numerical analysis as regards the low partial wave contribution to produce numerically exact results for the massive effective action. Comparing these against the results of the large mass expansion, the validity range of the large mass expansion is addressed. Also studied is the fermion mass dependence of the effective instanton-antiinstanton interaction.
\end{abstract}
\maketitle

\section{Introduction}
While in field theoretic studies we are often led to consider the one-loop effective action in some nontrivial backgrounds, it is quite difficult to have it explicitly evaluated. Also lacking are well-controlled approximation schemes for the quantity, which can cover broad types of backgrounds. This is true in four spacetime dimensions especially. Recently, for the case involving radially symmetric backgrounds, we (with G. Dunne) \cite{rea1,rea2} developed a new partial-wave-based calculational scheme, the partial wave cutoff method, by which the exact computation of fully renormalized one-loop effective actions can be performed explicitly. This method is a unique package of
analytical and numerical procedures (to treat high and low partial-wave contributions, respectively). So far it has been applied to the accurate determination of QCD single-instanton determinants for arbitrary quark mass values \cite{insdet}, to the prefactor calculation in the false vacuum decay \cite{falsevacuum}, and to the evaluation of the scalar one-loop effective actions (for any given mass values) in a class of Abelian or non-Abelian radially symmetric background gauge fields \cite{rea2}.
Also, very recently, the fermion one-loop effective action in Abelian radial background gauge fields has been studied by this method \cite{tba}, an important byproduct of this work being that there exist marked differences between the small mass mass limits of the derivative expansion for spinor and scalar theories.

In this paper the partial wave cutoff method will be used to study 4-D fermion one-loop effective actions in a class of genuinely \emph{non-Abelian}, radially symmetric, background gauge fields. This case differs from those of our earlier studies in that, as the differential operators pertaining to partial wave sectors are not completely separate, we here have to deal with an infinite number of functional determinants for \emph{matrix-valued radial differential operators}. One might then suspect that, because of technical difficulties in renormalizing the infinite product of such functional determinants and also in performing the needed numerical calculations, our whole approach becomes impractical in this case. Despite this complication, it will be demonstrated here that our method can suitably be extended such that the exact computation of the effective action becomes possible for this case as well. There are also issues specifically involving \emph{fermion} effective action (e.g., the massless limit behavior), and we intend to provide clarification on such aspect.

Specifically, in a 4-D Euclidean SU(2) gauge theory, we will consider in this work the one-loop effective action of a Dirac field (with mass $m$ and in the fundamental representation) when the background gauge fields are given as
\begin{eqnarray}
A_\mu (x) = \eta^{(\pm)}_{\mu\nu a} x_\nu f(r) \tau_a, \label{Amuform}
\end{eqnarray}
where $\mu,\nu=1,2,3,4$, $r\equiv \sqrt{x_\mu x_\mu}$, and $\eta^{(\pm)}_{\mu\nu a}$ are 't Hooft symbols \cite{thooft}. Writing $f(r)=\frac{1}{r^2}H(r)$, the radial function $H(r)$ is then assumed to have the form
\begin{eqnarray}
\text{(Case I)}&:& H_{\text{I}}(r;\alpha) = \frac{(r/\rho)^{2\alpha}}{1+(r/\rho)^{2\alpha}}, \qquad \left( |\alpha| \geq 1 \right), \label{case1} \\[6pt]
\text{(Case II)}&:& H_{\text{II}}(r;R,\beta) = \frac{(r/\rho)^2}{1+(r/\rho)^2} \frac{1+\tanh (\frac{r-R}{\beta \rho})}{2}, \quad (\beta \text{ can take \emph{either} sign}) \label{case2}
\end{eqnarray}
with constant background parameters $\rho$, $\alpha$, $R$ and $\beta$. This genuinely non-Abelian background field has been chosen so that one can learn something about the behavior of the corresponding fermion effective action as one changes the background parameters (and fermion mass $m$). Note that, in Case I (and Case II with $\beta>0$), our background gauge fields have the Pontryagin index equal to $\pm 1$; on the other hand, Case II with $\beta<0$ (and the ratio $R/\rho$ significantly larger than 1) corresponds to a well-separated instanton-antiinstanton configuration with zero Pontryagin index. [The Pontryagin index of the fields (\ref{Amuform}) is determined by the two numbers (see (\ref{pontryagin})), $H(0)$ and $H(\infty)$].
\begin{figure}
\subfigure[]{\includegraphics[scale=1.5]{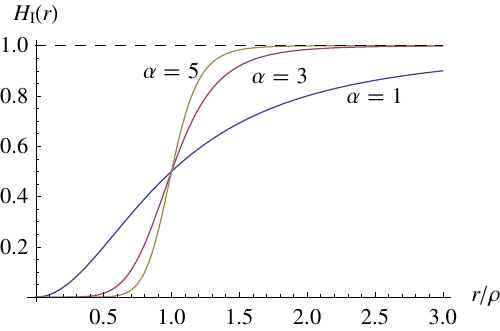}}
\subfigure[]{\includegraphics[scale=1.5]{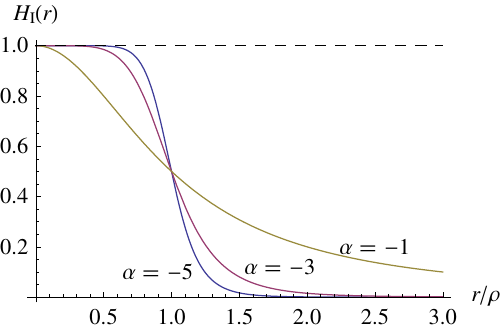}}
\subfigure[]{\includegraphics[scale=1.5]{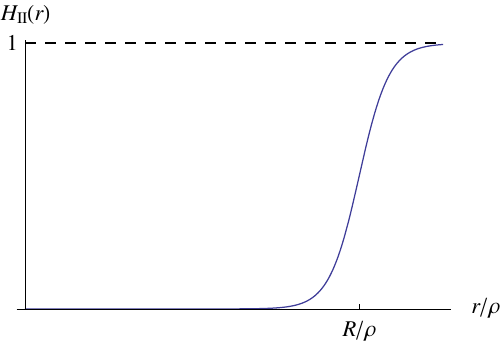}}
\subfigure[]{\includegraphics[scale=1.5]{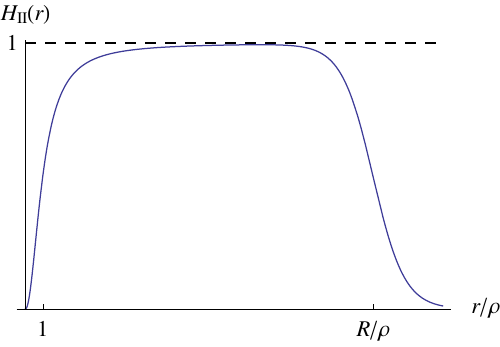}}
\caption{Plots of the radial profile function $H(r)$. For Case I, we have drawn the profile for $\alpha=1,3,5$ in (a) and $\alpha=-1,-3,-5$ in (b). Note that $H(r)$ behaves like a step function if $|\alpha|$ becomes very large. The plot in (c) --- a spherical-wall-like (anti-)instanton configuration --- is appropriate to Case II with $\beta>0$, and the plot in (d) --- an instanton-antiinstanton configuration --- corresponds to case II with $\beta<0$.} \label{fig1}
\end{figure}
In Fig. \ref{fig1} we have given the plots of the function $H(r)$ for some representative choices of our free parameters. Needless to say, with these backgrounds, the small mass limit of the fermion one-loop effective action becomes particularly interesting physically (because of the issue concerning fermion zero modes). In a well-separated instanton-antiinstanton configuration, the related effect
is also believed to generate
long-range interaction between instanton and antiinstanton \cite{callan,bardeen}. Our study should illuminate such aspect, too.

For the choice $\alpha=+1$ or $-1$ in Case I above, our fields (\ref{Amuform}) represent single instanton or antiinstanton \emph{solutions} \cite{bpst}, in the regular (for $\alpha=+1$) or singular (for $\alpha=-1$) gauge. These are (anti-)self-dual backgrounds, and here we have a simple relationship \cite{kwon} between the fermion one-loop effective action and that for a scalar field, which can be exploited for the effective action calculation. This was crucial in the calculation of \cite{insdet}. But, for other cases we will consider, we need to perform an entirely separate calculation for fermions; in this direct fermion analysis, additional complication due to the magnetic moment coupling term arises. Here we have found that the recently established chiral separation of the fermion effective action \cite{chirelpaper} can be utilized to the full advantage --- thanks to the latter, our technique to evaluate functional determinants of \emph{radially separable} differential operators can be extended to our problem involving Dirac fields.

This paper is organized as follows. In Sec. \ref{sec2} we describe a general outline for the calculational scheme we shall use, and also collect, for later use in the paper, various useful formulas. This is followed in Sec. \ref{sec3} by our analysis for the high partial-wave contribution to the fermion effective action; this part is calculated analytically using a WKB-type asymptotic series, for matrix-valued radial differential operators in this paper. We then combine this high partial-wave contribution with the low partial-wave contribution in Sec. \ref{sec4}, to obtain the explicit results for the renormalized fermion effective action with the background fields specified as above. An important part in the computation of the low partial wave contribution is the evaluation of functional determinants with matrix-valued differential operators, and for this we use the (generalized) Gel'fand-Yaglom approach \cite{gypaper}.
The partial wave contributions must be summed according to certain specific grouping procedure.
 In the limit the fermion mass approaches zero, we can give a semi-analytical treatment for partial-wave functional determinants and the results are used to discuss various features that the fermion effective action in the massless limit exhibits. (See for instance our formula (\ref{case1final}), where the exact small-mass-limit form of the fermion one-loop effective action in our general Case I backgrounds is given). This is possible because, unlike the scalar field case, partial-wave radial differential operators in the spinor case enjoys certain factorization property. Also presented are the results with nonzero mass. This involves an extensive numerical work, but, with our method of accelerating the convergence, numerically exact effective action can be obtained without excessive labor. For large fermion mass, we recover from our expression the result of large mass expansion. From our computation of the fermion effective action in Case II backgrounds, certain information on the effective instanton-antiinstanton interaction can also be gained. Sec. \ref{seccon} contains concluding remarks.

 Some supplementary results, related to the high partial-wave contribution of Sec. \ref{sec3}, can be found in Appendix \ref{appendixA}. In Appendix \ref{appendixB} we clarify some subtle aspect arising when one uses the Gel'fand-Yaglom approach in computing partial-wave functional determinants in the massless limit.

\section{Preparatory setup for our computation} \label{sec2}
\subsection{Fermion effective action in radial backgrounds}
The bare fermion effective action is
\begin{eqnarray}
\Gamma(A;m) &\sim& -\ln \det [-i \gamma\cdot D+m] + \text{const.} \nonumber\\[6pt]
&\sim& -\frac{1}{2} \ln \det [ (\gamma\cdot D)^2+m^2] + \text{const.},
\end{eqnarray}
where $\gamma\cdot D \equiv \gamma_\mu D_\mu$, $D_\mu = \partial_\mu -iA_\mu$, and $\{\gamma_\mu,\gamma_\nu\}=-2 \delta_{\mu\nu}$. Its Pauli-Villars regularized form, using the Schwinger proper-time representation \cite{schwinger}, is given by
\begin{eqnarray}
&& \Gamma_\Lambda(A;m) = \frac{1}{2} \int _0^{\infty }\frac{ds}{s} \left( e^{-m^2 s} - e^{-\Lambda^2 s} \right) F(s), \\[6pt]
&& F(s)=\int d^4x\; \mathrm{Tr} \left\langle x\left|\left[e^{-s (\gamma\cdot D)^2}-e^{-s \left(-\partial ^2\right)}\right]\right|x\right\rangle,
\end{eqnarray}
where `Tr' denotes the trace over Dirac spinor and internal isospin indices. Then the renormalized fermion effective action in the `minimal' subtraction scheme can be identified with the expression
\begin{eqnarray}
\Gamma_{\text{ren}}(A;m) = \lim_{\Lambda \to \infty} \left\{ \Gamma_\Lambda(A;m) - \frac{1}{3} \frac{1}{(4\pi)^2} \ln \frac{\Lambda^2}{\mu^2} \int d^4x\; \mathrm{tr} (F_{\mu\nu}F_{\mu\nu}) \right\}, \label{reneffectiveaction}
\end{eqnarray}
where $\mu$ is the renormalization scale, `tr' denotes the trace over isospin indices only, and $F_{\mu\nu}\equiv i[D_\mu,D_\nu]=\partial_\mu A_\nu - \partial_\nu A_\mu - i[A_\mu,A_\nu]$. For large mass $m$ the DeWitt WKB expansion (or heat kernel expansion) can be used to generate the large-mass approximate form for the effective action. But, in this paper, we are more interested in the \emph{exact} evaluation of the effective action in given backgrounds of the form (\ref{Amuform}) (but mass kept to arbitrary value).

To compute the effective action with the `radial' background (\ref{Amuform}), it is convenient to use the chiral representation for $\gamma$-matrices
\begin{eqnarray}
\gamma _{\mu }=\left(
\begin{array}{cc}
 0 & \sigma _{\mu } \\
 -\bar{\sigma }_{\mu } & 0
\end{array}
\right),\qquad (\text{with } \sigma _{\mu }=\left(\vec{\sigma},i\right) \text{ and } \bar{\sigma }_{\mu }=\left(\vec{\sigma},-i\right)=(\sigma_\mu^{\dagger})).
\end{eqnarray}
We then have
\begin{eqnarray}
\left(\gamma \cdot D \right){}^2=\left(
\begin{array}{cc}
 -D^2-\frac{1}{2} \eta _{\mu \nu a}^{(-)} \sigma _a F_{\mu \nu } & 0 \\
 0 & -D^2-\frac{1}{2} \eta _{\mu \nu a}^{(+)} \sigma _a F_{\mu \nu }
\end{array}
\right)
\end{eqnarray}
(here $D^2 \equiv D_\mu D_\mu$), and as a result the fermion effective action can be expressed by the sum of chirally projected ones \cite{chirelpaper}, viz.,
\begin{eqnarray}
\Gamma_{\text{ren}}(A;m) = \Gamma_{\text{ren}}^{(+)}(A;m) + \Gamma_{\text{ren}}^{(-)}(A;m)
\end{eqnarray}
with
\begin{eqnarray}
&& \Gamma _{\text{ren}}^{(\pm )}(A;m) = \lim_{\Lambda \to \infty } \frac{1}{2} \left\{\int _0^{\infty }\frac{ds}{s} \left(e^{-m^2 s}-e^{-\Lambda ^2 s}\right) F^{(\pm )}(s) \right. \nonumber\\
&&\qquad\qquad\qquad\qquad \left. -\frac{1}{3}\frac{1}{(4 \pi )^2} \ln \frac{\Lambda ^2}{\mu ^2} \int d^4x \left(\text{tr}\left(F_{\mu \nu } F_{\mu \nu }\right)\mp \frac{3}{2} \text{tr}\left(F_{\mu \nu }{}^* F_{\mu \nu }\right)\right)\right\}, \label{Gammarendef}\\[6pt]
&& F^{(\pm )}(s)=\int d^4x\; \overline{\mathrm{Tr}} \left\langle x\left|\left[e^{-s \left(-D^2-\frac{1}{2} \eta _{\mu \nu a}^{(\mp )} \sigma _a F_{\mu \nu }\right)}-e^{-s \left(-\partial ^2\right)}\right]\right|x\right\rangle, \label{Fdef}
\end{eqnarray}
where ${}^* F_{\mu \nu } \equiv \frac{1}{2} \epsilon_{\mu\nu\lambda\delta} F_{\lambda\delta}$ and `$\overline{\mathrm{Tr}}$' denotes the trace over 2-component (i.e., in the given chiral sector) spinor indices and isospin indices. Note that, to renormalize $\Gamma _{\text{ren}}^{(\pm )}$, we need to include the term proportional to the Pontryagin index in addition to the term involving the classical Yang-Mills action. To obtain the full effective action, there is no need to compute $\Gamma^{(+)}_{\text{ren}}$ and $\Gamma^{(-)}_{\text{ren}}$ separately, the two being related by \cite{chirelpaper}
\begin{eqnarray}
\Gamma _{\text{ren}}^{(+ )}(A;m) - \Gamma _{\text{ren}}^{(- )}(A;m) = \frac{1}{2} \frac{1}{(4\pi)^2} \ln \frac{m^2}{\mu^2} \int d^4x\; \mathrm{tr}(F_{\mu\nu}{}^* F_{\mu\nu}). \label{chiralrelation}
\end{eqnarray}
Explicit evaluation for one, the simpler from the two quantities $\Gamma _{\text{ren}}^{(+)}$ and $\Gamma _{\text{ren}}^{(- )}$ for a given background field, thus suffices. For our background fields (\ref{Amuform}), i.e., for $A_\mu(x)=\eta_{\mu\nu a}^{(\pm)} x_\nu f(r) \tau_a$, a simple calculation shows that $\eta_{\mu\nu a}^{(\pm)} \sigma_a F_{\mu\nu}$ (but \emph{not} $\eta_{\mu\nu a}^{(\mp)} \sigma_a F_{\mu\nu}$) takes a purely radial form
\begin{eqnarray}
\eta_{\mu\nu a}^{(\pm)} \sigma_a F_{\mu\nu} = -2 \left[4f(r)+r f'(r)-2 r^2f(r)^2 \right] \sigma_a \tau_a \equiv -2 g_F(r)  \sigma_a \tau_a, \label{gFdef}
\end{eqnarray}
i.e., a radially separable differential operator is obtained only for a particular chiral component of the fermion quadratic operator $(\gamma\cdot D)^2$. Hence, for our field $A_\mu(x)$ with the $\eta^{(+)}$-symbol chosen, it is the quantity $\Gamma_{\text{ren}}^{(-)}$ that we may try to evaluate directly by applying our partial wave cutoff method; for $\Gamma_{\text{ren}}^{(+)}$, on the other hand, we can use (\ref{chiralrelation}). So the full effective action $\Gamma_{\text{ren}}$ follows from the result for $\Gamma_{\text{ren}}^{(-)}$ alone. The situation is just the opposite if our field $A_\mu(x)$ happens to involve the $\eta^{(-)}$-symbol. [This choice of $\Gamma_{\text{ren}}^{(\pm)}$ depending on our background field form is appropriate with an \emph{arbitrary} radial function $f(r)$; for a particular form of $f(r)$ which gives rise to (anti-)self-dual field strengths, the other choice would be more suitable (with $\eta^{(\mp)}_{\mu\nu a} \sigma_a F_{\mu\nu} \equiv 0$ for the corresponding backgrounds)].

When the background field is given by the form (\ref{Amuform}), we have the classical action expressed using the function $H(r)=r^2f(r)$ as
\begin{eqnarray}
\frac{1}{2} \int d^4x\; \mathrm{tr} F_{\mu\nu}^2 = 12\pi^2 \int_0^\infty \frac{dr}{r} \left\{ r^2 H'(r)^2 + 4 H(r)^2 [H(r)-1]^2 \right\},
\end{eqnarray}
and the Pontryagin index as
\begin{eqnarray}
w=\frac{1}{16\pi^2} \int d^4x\; \mathrm{tr} F_{\mu\nu} {}^* F_{\mu\nu} = \mp \left. \left[ 2H(r)^3-3 H(r)^2 \right] \right|_{r=0}^{r=\infty}. \label{pontryagin}
\end{eqnarray}
Hence, if the $\eta^{(+)}$-symbol is chosen in our expression for $A_\mu(x)$, we find for the two Cases in (\ref{case1}) and (\ref{case2}) the Pontryagin index
\begin{eqnarray}
\text{(Case I)}&:& w= \left\{ \begin{array}{ccc}
 1&, & \alpha >1 \\
 -1&, & \alpha <-1
\end{array} \right., \\[6pt]
\text{(Case II)}&:& w= \left\{ \begin{array}{ccc}
 1&, & \beta >0 \\
 0&, & \beta <0
\end{array} \right. .
\end{eqnarray}
Also, for the background field form (\ref{Amuform}), the differential operator we must deal with, $-D^2-\frac{1}{2} \eta _{\mu \nu a}^{(\pm )} \sigma _a F_{\mu \nu }$, can be expressed in the form
\begin{eqnarray}
&& -D^2-\frac{1}{2} \eta _{\mu \nu a}^{(\pm)} \sigma _a F_{\mu \nu } \nonumber\\
&& \qquad =-\frac{\partial ^2}{\partial r^2}-\frac{3}{r} \frac{\partial }{\partial r}+\frac{4}{r^2} \vec{L}^2+8 f(r) \vec{T}\cdot\vec{L}^{(\pm)}+3 r^2 f(r)^2  +4 g_F(r) \vec{S}\cdot\vec{T}, \label{radialform1}
\end{eqnarray}
where $T^a \equiv \frac{1}{2} \tau_a$, $S^a \equiv \frac{1}{2} \sigma_a$, and $L^{(\pm)}_a$ and $\vec{L}^2$ are specified as \cite{thooft}
\begin{eqnarray}
L^{(\pm)}_a = -\frac{i}{2} \eta _{\mu \nu a}^{(\pm)} x_{\mu } \partial _{\nu }, \quad \left[L^{(\pm)}_a,L^{(\pm)}_b \right] = i \epsilon_{abc} L^{(\pm)}_c, \quad L^{(\pm)}_a L^{(\pm)}_a = \vec{L}^2.
\end{eqnarray}
For the evaluation of $\Gamma^{(\pm)}_{\text{ren}}(A;m)$, we may then resort to a kind of block diagonalization for the differential operator (\ref{radialform1}) in the form of partial waves. See Part B of this section..

Another useful information as regards our background field form (\ref{Amuform}) is that the appearance of the $\eta^{(+)}$- or $\eta^{(-)}$-symbol in the expression is actually tied up with the gauge choice. Explicitly, using the relation
\begin{eqnarray}
\eta _{\mu \nu a}^{(\pm)} \frac{x_\nu}{r^2} \tau^a = i\Omega_{(\pm)}^{-1}(x) \partial_\mu \Omega_{(\pm)} (x), \quad \left(\Omega_{(\pm)}(x) = \frac{x_4\mp i \vec{x}\cdot\vec{\tau}}{r} \in \mathrm{SU}(2)\right)
\end{eqnarray}
it is not difficult to show that
\begin{eqnarray}
\Omega_{(\pm)}(x) \left[ \eta _{\mu \nu a}^{(\pm)} \frac{x_\nu}{r^2} H(r) \tau_a \right] \Omega_{(\pm)}^{-1}(x) + i\Omega_{(\pm)}(x) \partial_\mu \Omega_{(\pm)}^{-1} (x) = \eta _{\mu \nu a}^{(\mp)} \frac{x_\nu}{r^2} [1-H(r)] \tau_a,
\end{eqnarray}
i.e., under the (singular) gauge transformation involving the SU(2) matrix $\Omega_{(\pm)}(x)$, our $\eta^{(\pm)}$-symbol form with the radial function $H(r)$ goes over to the $\eta^{(\mp)}$-symbol form with the radial function $1-H(r)$. This has the consequence that an identical fermion effective action $\Gamma_{\text{ren}}(A;m)$ would result either with the form $A_\mu(x) = \eta^{(\pm)}_{\mu\nu a} \frac{x_\nu}{r^2} H(r) \tau_a$ or with the form $\bar{A}_\mu(x) = \eta^{(\mp)}_{\mu\nu a} \frac{x_\nu}{r^2} \bar{H}(r) \tau_a$ where $\bar{H}(r) \equiv 1-H(r)$. Then notice that, for $H(r)$ given by the form in (\ref{case1}), we find
\begin{eqnarray}
\bar{H}_{\text{I}}(r) = 1 - \frac{(r/\rho)^{2 \alpha}}{1+(r/\rho)^{2 \alpha}} = \frac{(r/\rho)^{-2 \alpha}}{1+(r/\rho)^{-2 \alpha}}.
\end{eqnarray}
Therefore, in our Case I we may just evaluate the effective action with the radial function of the background field chosen as
\begin{eqnarray}
H_{\text{I}}(r;\alpha) = \frac{(r/\rho)^{2 \alpha}}{1+(r/\rho)^{2 \alpha}}, \qquad (\alpha \geq 1); \label{chosenH1}
\end{eqnarray}
then the gauge invariance of the effective action tells us the result appropriate to the same function form for $H(r)$ but with $\alpha \leq -1$. Also, applying the above gauge transformation to our Case II, we may replace our radial profile function in (\ref{case2}) by the form
\begin{eqnarray}
\bar{H}_{\text{II}}(r;R,\beta) &=& 1 - \frac{(r/\rho)^2}{1+(r/\rho)^2} + \frac{(r/\rho)^2}{1+(r/\rho)^2} \frac{1-\tanh (\frac{r-R}{\beta \rho})}{2} \nonumber\\[6pt]
&=& \frac{(r/\rho)^{-2}}{1+(r/\rho)^{-2}} + H_{\text{II}}(r;R,-\beta), \label{HbarII}
\end{eqnarray}
as this form should lead to the same set of effective actions. Note that for $\beta<0$ and $R$ large, what we have in (\ref{HbarII}) is the sum of a singular-gauge instanton (antiinstanton) located near the origin and a spherical-wall-like antiinstanton (instanton) configuration at large radius $R$.
\begin{figure}
\subfigure[]{\includegraphics[scale=1.5]{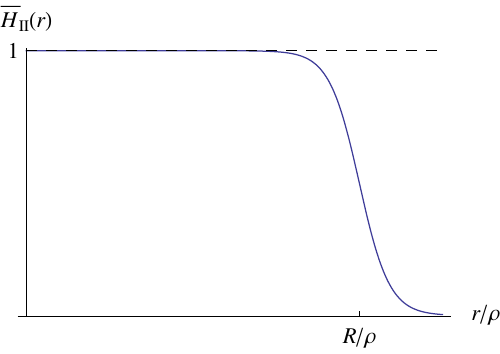}}
\subfigure[]{\includegraphics[scale=1.5]{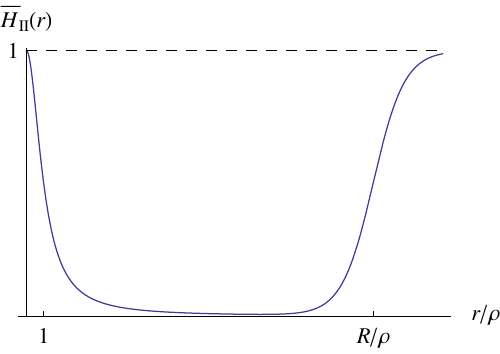}}
\caption{Plots of the radial profile function $\bar{H}_{\text{II}}(r)$ given in (\ref{HbarII}): (a) for $\beta>0$ and (b) for $\beta<0$.} \label{fig2}
\end{figure}
See Fig \ref{fig2} for the illustration of the radial profile function corresponding to our form (\ref{HbarII}).

\subsection{Partial-wave decomposed form}
The differential operator in (\ref{radialform1}) can be decomposed into an infinite number of partial-wave radial differential operators (with matrix coefficients). For our partial waves, let us consider the basis $|j,j_3,q,l,\bar{l}_3\rangle$ where various quantum numbers introduced are specified by
\begin{eqnarray}
(\vec{L}^2)' &=& l(l+1), \qquad\qquad l=0,\frac{1}{2},1,\frac{3}{2},\cdots; \nonumber\\
(\vec{Q}^2)' &=& q(q+1), \qquad\qquad (\text{with } Q_a \equiv L_a^{(\pm)}+T_a),\; q=\left| l\pm \frac{1}{2} \right|; \nonumber\\
(\vec{J}^2)' &=& j(j+1), \qquad\qquad (\text{with } J_a \equiv Q_a+S_a),\; j=\left| q\pm \frac{1}{2} \right|; \label{angnumdef}\\
(J_3)' &=& j_3 = -j,-j+1,\cdots,j; \nonumber\\
(L_3^{(\pm)})' &=& \bar{l}_3 = -l,-l+1,\cdots,l. \nonumber
\end{eqnarray}
In this basis, the operator in (\ref{radialform1}) is not completely diagonal, but we may still write it (for given values of $l$ and $j$) as
\begin{eqnarray}
-D^2-\frac{1}{2} \eta _{\mu \nu a}^{(\pm)} \sigma _a F_{\mu \nu } \longrightarrow \mathcal{H}_{l,j}&=&-\frac{\partial ^2}{\partial r^2}-\frac{3}{r} \frac{\partial }{\partial r}+\frac{4l(l+1)}{r^2}+4 f(r) \left[ q(q+1)-l(l+1)-\frac{3}{4} \right] \nonumber\\[6pt]
&& +3 r^2 f(r)^2 +4 g_F(r) \vec{S}\cdot\vec{T} \nonumber\\[6pt]
&\equiv& -\partial_{(l)}^2 + \mathcal{V}_{l,j}(r) \label{Hform}
\end{eqnarray}
with suitable matrix $\vec{S}\cdot\vec{T}$ in the space of allowed $q$-states (for given $l$, $j$). [In (\ref{Hform}), $\partial_{(l)}^2 = \frac{\partial ^2}{\partial r^2}+\frac{3}{r} \frac{\partial }{\partial r}-\frac{4l(l+1)}{r^2}$ represents the 4-D Laplacian $\partial_\mu\partial_\mu$ for given angular momentum]. As for the matrix $\vec{S}\cdot\vec{T}$, we here note that, for given $l$, the quantum number $q$ should be equal to $l+\frac{1}{2}$ when $j=l+1$ and equal to $l-\frac{1}{2}$ when $j=l-1$, while $q$ can take either value of $l\pm\frac{1}{2}$ when $j=l\neq 0$. When $j=l=0$, only the value $q=\frac{1}{2}$ is available. Then, after somewhat lengthy but straightforward calculations, we obtain following representations for $\vec{S}\cdot\vec{T}$:
\begin{eqnarray}
\begin{array}{lcl}
\vec{S}\cdot\vec{T} \longrightarrow \displaystyle{\frac{1}{4}} &,& \text{if } j=l\pm1 \; \left(\text{and } q=l\pm\displaystyle{\frac{1}{2}} \right),\\[6pt]
\vec{S}\cdot\vec{T} \longrightarrow \displaystyle{\frac{1}{4(2l+1)}} \left(
\begin{array}{cc}
 -2l-3 & 4\sqrt{l (l+1)} \\
 4\sqrt{l (l+1)} & -2 l+1
\end{array}
\right)&,& \text{if } j=l\neq 0, \\[6pt]
\vec{S}\cdot\vec{T} \longrightarrow -\displaystyle{\frac{3}{4}}&,& \text{if } j=l=0 \; \left(\text{and } q=\displaystyle{\frac{1}{2}} \right),
\end{array} \label{STform}
\end{eqnarray}
where our $2\times 2$ matrix form for $j=l\neq 0$ is written relative to the basis $(|q=l+\frac{1}{2}\rangle, |q=l-\frac{1}{2}\rangle)$. Using the expression in (\ref{STform}), we thus find, for the `potential' $\mathcal{V}_{l,j}(r)$ defined in (\ref{Hform}),
\begin{eqnarray}
&& \mathcal{V}_{l,l+1}(r) = 3 r^2 f(r)^2+4 l f(r)+g_F(r), \label{Vform1}\\[6pt]
&& \mathcal{V}_{l,l-1}(r) = 3 r^2 f(r)^2-4 (l+1) f(r)+g_F(r), \\[6pt]
&& \mbox{\boldmath$\mathcal{V}$}_{l,l}(r) = 3 r^2 f(r)^2 +4 f(r) \left(
\begin{array}{cc}
 l & 0 \\
 0 & -l-1
\end{array}
\right)+\frac{g_F(r)}{2l+1} \left(
\begin{array}{cc}
 -2l-3 & 4\sqrt{l (l+1)} \\
 4\sqrt{l (l+1)} & -2 l+1
\end{array}
\right), \nonumber\\[6pt]
&& \qquad\qquad\qquad\qquad\qquad\qquad\qquad\qquad\qquad\qquad\qquad\qquad\qquad\qquad (l\neq0) \label{matrixV}
\end{eqnarray}
with $g_F(r) \equiv 4f(r)+rf'(r)-2r^2f(r)^2$ (see (\ref{gFdef})), and
\begin{eqnarray}
\mathcal{V}_{0,0} = 3 r^2 f(r)^2-3 g_F(r). \qquad\qquad\qquad\qquad\qquad\qquad\qquad \label{Vform2}
\end{eqnarray}
[The bold-faced letter for $\mbox{\boldmath$\mathcal{V}$}_{l,l}(r)$ is to indicate that it is matrix-valued]. To each partial wave labeled by quantum numbers $J\equiv (l,j,j_3,\bar{l}_3)$ corresponds the (matrix) radial differential operator $\mathcal{H}_{l,j}$ specified by (\ref{Hform}) and (\ref{Vform1})-(\ref{Vform2}).

Our formula (\ref{Gammarendef}) for the effective action $\Gamma_{\text{ren}}^{(\pm)}(A;m)$ can then be reexpressed using the quantity involving the radial differential operator $\mathcal{H}_{l,j}$ appropriate to partial waves. Here note that, as should be evident from the discussion above, $\Gamma^{(+)}_{\text{ren}}(A;m)$ in the background (\ref{Amuform}) with the $\eta^{(-)}$-symbol picked has the same value as $\Gamma^{(-)}_{\text{ren}}(A;m)$ in the very background but with the $\eta^{(+)}$-symbol taken. Then, thanks to (\ref{chiralrelation}) and (\ref{pontryagin}), we are led to conclude that the full effective action $\Gamma_{\text{ren}}(A;m)$ in the background (\ref{Amuform}) is represented by the same function of $f(r)$ irrespectively of which $\eta$-symbol is picked. Knowing this, we may well consider only the background field form
\begin{eqnarray}
A_\mu (x) = \eta^{(+)}_{\mu\nu a} x_\nu f(r) \tau_a, \qquad \left( f(r) \equiv \frac{1}{r^2} H(r) \right) \label{Amuform2}
\end{eqnarray}
from now on and go on to evaluate the quantity $\Gamma^{(-)}_{\text{ren}}(A;m)$, a particular chiral projection of $\Gamma_{\text{ren}}(A;m)$, in this background. The radial function $H(r)$ is that of the form (\ref{chosenH1}) or of (\ref{case2}). Note that, because of (\ref{chiralrelation}), the full renormalized effective action can then be found simply by using the formula
\begin{eqnarray}
\Gamma_{\text{ren}}(A;m) = 2 \Gamma_{\text{ren}}^{(-)}(A;m) + \frac{1}{2} \frac{1}{(4\pi)^2} \ln \frac{m^2}{\mu^2} \int d^4x\; \mathrm{tr}(F_{\mu\nu}{}^* F_{\mu\nu}). \label{chiralformula}
\end{eqnarray}
Now, for the partial-wave-decomposed form of $\Gamma^{(-)}_{\text{ren}}(A;m)$, we may express the function $F^{(-)}(s)$ (see (\ref{Fdef})) by the form
\begin{eqnarray}
F^{(-)}(s) &=& \sum_J F^{(-)}_{l,j}(s), \nonumber\\[6pt]
F^{(-)}_{l,j}(s) &=& \int_0^\infty dr\; \mathrm{tr}\! \left\{ G_{l,j}(r,r;s) - G_l^{\text{free}}(r,r;s) \right\}
\end{eqnarray}
where we introduced the radial proper-time Green's functions
\begin{eqnarray}
G_{l,j}(r,r';s) \equiv \left\langle r \left| e^{-s \tilde{\mathcal{H}}_{l,j}} \right| r' \right\rangle, \quad G_l^{\text{free}}(r,r';s) \equiv \left\langle r \left| e^{-s (-\tilde{\partial}_{(l)}^2)} \right| r' \right\rangle
\end{eqnarray}
of the operators $\tilde{\mathcal{H}}_{l,j}$ and $\tilde{\partial}_{(l)}^2$ defined through \cite{rea1}
\begin{eqnarray}
\tilde{\mathcal{H}}_{l,j} &\equiv& \frac{1}{r^{3/2}} \mathcal{H}_{l,j} r^{3/2} \nonumber\\[6pt]
&=& -\frac{d^2}{dr^2} + \frac{4l(l+1)+\frac{3}{4}}{r^2} +\mathcal{V}_{l,j}(r) \; \left( \equiv -\tilde{\partial}_{(l)}^2 + \mathcal{V}_{l,j}(r) \right).
\end{eqnarray}

When $l$ is large, we can employ the WKB method to calculate $F^{(-)}_{l,j}(s)$ systematically; this, we do in Sec. \ref{sec3}. (Using the result, the renormalization problem is also solved in an expedient way). But we need a different strategy for small $l$, and to determine the related contribution to $\Gamma^{(-)}_{\text{ren}}(A;m)$ it is more convenient to consider the quantity resulting after $s$-integration,
\begin{eqnarray}
\int_0^\infty \frac{ds}{s} e^{-m^2s} F^{(-)}_{l,j}(s) = \ln \!\left[ \frac{\det(\tilde{\mathcal{H}}_{l,j}+m^2)}{\det(-\tilde{\partial}_{(l)}^2+m^2)}\right] = \ln \!\left[ \frac{\det(\mathcal{H}_{l,j}+m^2)}{\det(-\partial_{(l)}^2+m^2)}\right].
\end{eqnarray}
Then, given a specific background, we may evaluate the latter quantity with the help of the Gel'fand-Yaglom method. We do this for our (matrix) differential operators $\mathcal{H}_{l,j}$ in Sec. \ref{sec4}. In performing these calculations, some care must be exercised if the effective radial potential
\begin{eqnarray}
V_{l,j}(r) = \frac{4l(l+1)+\frac{3}{4}}{r^2} + \mathcal{V}_{l,j}(r) \label{Vdef}
\end{eqnarray}
doest not have the same small-$r$ and large-$r$ behavior as $V_l^{\text{free}}(r) = \frac{4l(l+1)+3/4}{r^2}$.
In the latter case,
to avoid the appearance of largely oscillating terms in the process of summing partial wave contributions,
a certain grouping of terms might be contemplated \cite{insdet}; this is relevant also with our background fields since,
or our Case I for instance, we have $f(r)\sim\frac{1}{r^2}$ and so $V_{l,j}(r)\sim \frac{4q(q+1)+3/4}{r^2}$ as $r\to\infty$.

To facilitate our discussions, we will now write out the mathematical expressions we must evaluate to determine $\Gamma^{(-)}_{\text{ren}}(A;m)$ for our background fields. Following Refs. \cite{insdet}, we introduce the (floating) parameter $L$ as the partial wave cutoff and write
\begin{eqnarray}
\Gamma^{(-)}_{\text{ren}}(A;m) = \lim_{L\to\infty} \left[ \Gamma^{(-)}_{l\leq L}(A;m) +\Gamma^{(-)}_{l> L}(A;m) \right]; \label{devideGamma}
\end{eqnarray}
here, $\Gamma^{(-)}_{l\leq L}(A;m)$ represents the `low' partial wave contribution given as
\begin{eqnarray}
&& \Gamma^{(-)}_{l\leq L}(A;m) = -\frac{1}{2} \left( \left\{ \ln \!\left[ \frac{\det(\mathcal{H}_{0,0}+m^2)}{\det(-\partial_{(l=0)}^2+m^2)}\right] - \ln \!\left[ \frac{\det(\mathcal{H}_{0,1}+m^2)}{\det(-\partial_{(l=0)}^2+m^2)}\right] \right\} \right. \nonumber\\[6pt]
&& +\sum_{l=\frac{1}{2},1,\cdots}^L \left\{ (2l+1)^2 \left( \ln \!\left[ \frac{\det(\mbox{\boldmath$\mathcal{H}$}_{l,l}+m^2)}{\left\{\det(-\partial_{(l)}^2+m^2)\right\}^2}\right] + \ln \!\left[ \frac{\det(\mathcal{H}_{l-\frac{1}{2},l+\frac{1}{2}}+m^2)}{\det(-\partial_{(l-\frac{1}{2})}^2+m^2)}\right] \right.\right. \label{GammaL}\\[6pt]
&& \left.\left.\left. +\ln \!\left[ \frac{\det(\mathcal{H}_{l+\frac{1}{2},l-\frac{1}{2}}+m^2)}{\det(-\partial_{(l+\frac{1}{2})}^2+m^2)}\right] \right) -\left( \ln \!\left[ \frac{\det(\mathcal{H}_{l,l+1}+m^2)}{\det(-\partial_{(l)}^2+m^2)}\right] + \ln \!\left[ \frac{\det(\mathcal{H}_{l+\frac{1}{2},l-\frac{1}{2}}+m^2)}{\det(-\partial_{(l+\frac{1}{2})}^2+m^2)}\right] \right)\right\}\right), \nonumber
\end{eqnarray}
while $\Gamma^{(-)}_{l> L}(A;m)$, the `high' partial wave contribution, can be expressed by the form
\begin{eqnarray}
\Gamma^{(-)}_{l> L}(A;m) &=& \lim_{\Lambda \to \infty } \frac{1}{2} \left\{ \int _0^{\infty }\frac{ds}{s} \left(e^{-m^2 s}-e^{-\Lambda ^2 s}\right) \int_0^\infty dr\; \sum_{l=L+\frac{1}{2}}^\infty \mathcal{G}_l(r,r;s) \right. \nonumber\\[6pt]
&& \qquad \left. -\frac{1}{3}\frac{1}{(4 \pi )^2} \ln \frac{\Lambda ^2}{\mu ^2} \int d^4x \left(\text{tr}\left(F_{\mu \nu } F_{\mu \nu }\right)+ \frac{3}{2} \text{tr}\left(F_{\mu \nu }{}^* F_{\mu \nu }\right)\right)\right\}, \label{GammaH}
\end{eqnarray}
where we defined
\begin{eqnarray}
&& \mathcal{G}_l(r,r;s) = (2l+1)^2 \left\{ \mathrm{tr}\, \mathbf{G}_{l,l}(r,r;s) + G_{l-\frac{1}{2},l+\frac{1}{2}}(r,r;s) + G_{l+\frac{1}{2},l-\frac{1}{2}}(r,r;s) \right. \nonumber\\
&& \qquad \left. - 2 G_l^{\text{free}}(r,r;s) - G_{l+\frac{1}{2}}^{\text{free}}(r,r;s) - G_{l-\frac{1}{2}}^{\text{free}}(r,r;s) \right\} -\Big\{ G_{l,l+1}(r,r;s)  \nonumber\\
&& \qquad  + G_{l+\frac{1}{2},l-\frac{1}{2}}(r,r;s) - G_l^{\text{free}}(r,r;s) - G_{l+\frac{1}{2}}^{\text{free}}(r,r;s) \Big\}, \quad \left(l=L+\frac{1}{2},L+1,\cdots\right). \qquad \label{calGdef}
\end{eqnarray}
In (\ref{GammaL}) and (\ref{GammaH}), various partial wave contributions have been grouped
(with correct degeneracy factors)
in such a way that we may have both the small-$r$ and large-$r$ behaviors of the effective potential matched when the radial differential operators figuring in within each group are taken together. (This grouping is especially important when we discuss the massless limit of the effective action (see Part A of Sec. \ref{sec4})).
In Case II background with $\beta<0$ (and {\em finite} $R$), the above partial-wave grouping is not really required;
but it is desirable to have a procedure applicable to all cases we will consider.

We remark that (\ref{devideGamma}), even without taking the limit $L\to\infty$, corresponds to an exact relation. But, only by taking $L$ to be relatively large, this relation can be put to use as a powerful calculational tool for the effective action --- this is because, for large $L$, two independent means to determine separately the high and low partial wave contributions become available. This is the key element of the partial-wave cutoff method. In subsequent sections, we shall obtain the expression for $\Gamma^{(-)}_{l> L}(A;m)$ as a $\frac{1}{L}$-series form analytically while the other piece, $\Gamma^{(-)}_{l\leq L}(A;m)$, in our chosen background fields will be found numerically (and semi-analytically in the small mass limit). If the sum of the two contributions yields an $L$-independent result, it is of course the sign that we have secured exact result for $\Gamma^{(-)}_{\text{ren}}(A;m)$.

\section{High partial-wave contribution to the effective action} \label{sec3}
In this section we shall calculate the large-$L$ form of the expression (\ref{GammaH}) to desired accuracy. To that end we need a systematic large-$l$ approximation to the function $\mathcal{G}_l(r,r;s)$, which is valid uniformly for $s$ in the range $0<sl^2<O(1)$ \cite{rea1}. For this one might consider using the WKB approximation \cite{insdet,insplb}; but, with a matrix-valued potential, it is not trivial to apply the WKB method directly. Alternative methods, yet serving our purpose in a satisfactory way, were found in \cite{rea1,rea3}. These latter approaches are not only simpler but also easily extendable to the case involving matrix-valued potentials. In this paper we will specifically use the method of \cite{rea3} to generate the desired $\frac{1}{l}$-expansion of $G_{l,j}(r,r;s)$, because of its convenience in dealing with a matrix-valued potential.

The idea of \cite{rea3} is to rewrite the proper-time Green function, introducing the momentum-like variable $p$, in the form
\begin{eqnarray}
G_{l,j}\left(r,r';s\right) &=& \int _{-\infty }^{\infty }\frac{dp}{2 \pi } \left\langle r\left|e^{-s \tilde{\mathcal{H}}_{l,j}}\right|p\right\rangle \left\langle p\left|r'\right.\right\rangle \nonumber\\[6pt]
&=& \int _{-\infty }^{\infty }\frac{dp}{2 \pi } e^{-s \left[-\partial_r^2 +V_{l,j}(r)\right]} e^{-i p \left(r-r'\right)}, \label{Gexpand}
\end{eqnarray}
and to find a convenient way to study directly the $r'=r$ limit of this function. Then, after moving the last Fourier factor $e^{-i p \left(r-r'\right)}$ in (\ref{Gexpand}) to the left of the differential operator $\partial_r(\equiv \frac{\partial}{\partial r})$, we may take the coincidence limit $r'=r$ to obtain the following representation:
\begin{eqnarray}
G_{l,j}\left(r,r;s\right) &=& \int _{-\infty }^{\infty }\frac{dp}{2 \pi } e^{-s \left[-(\partial_r-ip)^2 +V_{l,j}(r)\right]} \cdot 1 \nonumber\\[6pt]
&\equiv& \int _{-\infty }^{\infty }\frac{dp}{2 \pi } K(r,p;s). \label{GfromK}
\end{eqnarray}
The function $K(r,p;s)$ introduced here can be identified with the solution of the differential equation
\begin{eqnarray}
\left\{\frac{\partial }{\partial s}-\left(\partial_r-ip \right)^2 + V_{l,j}(r)\right\} K(r,p;s)=0 \label{Keq}
\end{eqnarray}
under the boundary condition
\begin{eqnarray}
K(r,p;s=0) =1. \label{Kboundarycondition}
\end{eqnarray}

Because of the connection (\ref{GfromK}), the desired large-$l$ series for $G_{l,j}(r,r;s)$ will follow immediately if we have an appropriate development for the function $K(r,p;s)$. For our investigation, it is convenient to regard rescaled variables $t=sl^2$ and $q=\frac{p}{l}$ (rather than $s$ and $p$) as independent variables of the function of $K$. Also we express the potential $V_{l,j}(r)$ (see (\ref{Vdef})) as a series in $\frac{1}{l}$, i.e.,
\begin{eqnarray}
V_{l,j}(r) = l^2 \frac{4}{r^2} + l\, \bar{V}_1(r) + \bar{V}_2(r) + \frac{1}{l}\bar{V}_3(r) + \frac{1}{l^2}\bar{V}_4(r) + \cdots, \label{Vseries}
\end{eqnarray}
where $\bar{V}_k(r)$ are some $l$-independent (matrix) functions. Then (\ref{Keq}) can be written in the form
\begin{eqnarray}
\left\{\frac{\partial }{\partial t}-\left(\frac{1}{l} \frac{\partial }{\partial r}-i q\right)^2+\frac{4}{r^2}+\frac{\bar{V}_1(r)}{l}+\frac{\bar{V}_2(r)}{l^2}+\cdots \right\} K\!\left(r,q l;\frac{t}{l^2}\right)=0. \label{Ksereq}
\end{eqnarray}
This equation may be solved by positing the ansatz
\begin{eqnarray}
K\!\left(r,q l;\frac{t}{l^2}\right)=K_0(r,q,t) \left\{1+\frac{a_1(r,q,t)}{l}+\frac{a_2(r,q,t)}{l^2}+\cdots \right\}, \label{Kseries}
\end{eqnarray}
where the function $K_0(r,q,t)$, the $l\to\infty$ form of $K(r,q l;\frac{t}{l^2})$, is chosen to satisfy
\begin{eqnarray}
\left(\frac{\partial }{\partial t}+q^2+\frac{4}{r^2}\right) K_0(r,q,t)=0. \label{K0eq}
\end{eqnarray}
The solution to (\ref{K0eq}), subject to the boundary condition (\ref{Kboundarycondition}), is
\begin{eqnarray}
K_0(r,q,t)=e^{-t \left(q^2+\frac{4}{r^2}\right)}. \label{K0}
\end{eqnarray}
To determine $a_k(r,q,t)$ ($k=1,2,\cdots$), we may plug in (\ref{Kseries}) and (\ref{K0}) into (\ref{Ksereq}) to obtain following recurrence relations for them:
\begin{eqnarray}
\frac{\partial a_k}{\partial t}=-2 i q \left(\frac{\partial }{\partial r}+\frac{8 t}{r^3}\right) a_{k-1}+\left(\frac{\partial }{\partial r}+\frac{8 t}{r^3}\right)^2 a_{k-2}-\sum _{n=1}^k \bar{V}_n a_{k-n}, \quad (k=1,2,\cdots)
\end{eqnarray}
with $a_0=1$ and $a_{-1}=0$. Then we can find $a_k$'s successively (with $a_k(r,q,t=0)=0$ for all $k=1,2,\cdots$): here, the first few terms which are indispensable for our calculation using (\ref{devideGamma}) are
\begin{eqnarray}
&& a_1 = -t \bar{V}_1-\frac{8 i q t^2}{r^3}, \\[6pt]
&& a_2 = \frac{8 i q t^3}{r^3} \bar{V}_1+i q t^2 \bar{V}_1'+\frac{t^2}{2} \bar{V}_1^2-t \bar{V}_2 -\frac{32 q^2 t^4}{r^6}+\frac{16 q^2 t^3}{r^4}+\frac{64 t^3}{3 r^6}-\frac{12 t^2}{r^4}, \\[6pt]
&& a_3 = \left(\frac{8 q^2 t^4}{r^3}-\frac{16 t^3}{3 r^3}\right)\bar{V}_1'+ \left(\frac{32 q^2 t^5}{r^6}-\frac{16 q^2 t^4}{r^4}-\frac{64 t^4}{3 r^6}+\frac{12 t^3}{r^4}\right)\bar{V}_1+\left(\frac{2 q^2 t^3}{3}-\frac{t^2}{2}\right) \bar{V}_1'' \nonumber\\
&& \quad -\frac{4 i q t^4}{r^3} \bar{V}_1^2+\frac{8 i q t^3}{r^3} \bar{V}_2-\frac{i q t^3}{3} \left( 2 \bar{V}_1 \bar{V}_1'+ \bar{V}_1' \bar{V}_1 \right)+i q t^2 \bar{V}_2'-\frac{t^3}{6} \bar{V}_1^3+\frac{t^2}{2} \left(\bar{V}_1 \bar{V}_2+\bar{V}_2 \bar{V}_1\right) \nonumber\\
&& \quad -t \bar{V}_3+\frac{256 i q^3 t^6}{3 r^9}-\frac{128 i q^3 t^5}{r^7}+\frac{32 i q^3 t^4}{r^5}-\frac{512 i q t^5}{3 r^9}+\frac{256 i q t^4}{r^7}-\frac{64 i q t^3}{r^5}, \\[6pt]
&& a_4 = \frac{512 q^4 t^8}{3 r^{12}}-\frac{512 q^4 t^7}{r^{10}}-\frac{2048 q^2 t^7}{3 r^{12}}+\frac{384 q^4 t^6}{r^8}+\frac{6016 q^2 t^6}{3 r^{10}}+\frac{2048 t^6}{9 r^{12}}+\frac{4 i q t^5}{3 r^3} \bar{V}_1^3 \nonumber\\
&& \quad +\left( \frac{8 q^2 t^5}{r^4}+\frac{32 t^5}{3 r^6}-\frac{6 t^4}{r^4}-\frac{16 q^2 t^6}{r^6} \right) \bar{V}_1^2 -\frac{8 q^2 t^5}{3 r^3} \left( 2 \bar{V}_1 \bar{V}_1'+\bar{V}_1' \bar{V}_1\right)-\frac{64 q^4 t^5}{r^6}-\frac{1472 q^2 t^5}{r^8} \nonumber\\
&& \quad -\frac{3328 t^5}{5 r^{10}}+\left(\frac{t^3}{3}-\frac{q^2 t^4}{2} \right) \left(\bar{V}_1'\right)^2 -\frac{4 i q t^4}{r^3} \left(\bar{V}_1 \bar{V}_2+ \bar{V}_2 \bar{V}_1\right) +\frac{t^4}{r^3} \left( \frac{10}{3} \bar{V}_1 \bar{V}_1' +2 \bar{V}_1' \bar{V}_1\right) \nonumber\\
&& \quad -q^2t^4 \left( \frac{\bar{V}_1 \bar{V}_1''}{2}+\frac{\bar{V}_1'' \bar{V}_1}{6} \right)+iqt^4 \left( \frac{\bar{V}_1^2 \bar{V}_1'}{4}+\frac{\bar{V}_1' \bar{V}_1^2}{12}+\frac{\bar{V}_1 \bar{V}_1' \bar{V}_1}{6} \right) +\frac{t^4}{24} \bar{V}_1^4+\frac{240 q^2 t^4}{r^6}+\frac{488 t^4}{r^8} \nonumber\\
&& \quad -\frac{i q t^3}{3} \left( 2 \bar{V}_1 \bar{V}_2'+ 2 \bar{V}_2 \bar{V}_1'+\bar{V}_1' \bar{V}_2+\bar{V}_2' \bar{V}_1 \right)+t^3 \left( \frac{\bar{V}_1 \bar{V}_1''}{3}+\frac{\bar{V}_1'' \bar{V}_1}{6} \right)+\frac{8 i q t^3}{r^3} \bar{V}_3-\frac{80 t^3}{r^6} \nonumber\\
&& \quad -\frac{t^3}{6} \left( \bar{V}_1^2 \bar{V}_2+ \bar{V}_2 \bar{V}_1^2+\bar{V}_1 \bar{V}_2 \bar{V}_1 \right)+\frac{t^2}{2} \left( \bar{V}_2^2+\bar{V}_1 \bar{V}_3+\bar{V}_3 \bar{V}_1 \right)-t \bar{V}_4+\left(\frac{2 q^2 t^3}{3}-\frac{t^2}{2}\right) \bar{V}_2'' \nonumber\\
&& \quad +\left(\frac{64 i q t^4}{r^5}-\frac{256 i q^3 t^7}{3 r^9}+\frac{128 i q^3 t^6}{r^7}+\frac{512 i q t^6}{3 r^9}-\frac{32 i q^3 t^5}{r^5}-\frac{256 i q t^5}{r^7}\right) \bar{V}_1+i q t^2 \bar{V}_3' \nonumber\\
&& \quad +\left(\frac{32 q^2 t^5}{r^6}-\frac{16 q^2 t^4}{r^4}-\frac{64 t^4}{3 r^6}+\frac{12 t^3}{r^4}\right) \bar{V}_2+\left(\frac{16 i q^3 t^5}{r^4}-\frac{32 i q^3 t^6}{r^6}+\frac{64 i q t^5}{r^6}-\frac{32 i q t^4}{r^4}\right) \bar{V}_1' \nonumber\\
&& \quad +\left(\frac{8 q^2 t^4}{r^3}-\frac{16 t^3}{3 r^3}\right) \bar{V}_2'+\left(\frac{32 i q t^4}{3 r^3}-\frac{16 i q^3 t^5}{3 r^3}\right) \bar{V}_1''+\left(\frac{2 i q t^3}{3}-\frac{ i q^3 t^4}{3}\right) \bar{V}_1^{(3)},
\end{eqnarray}
where $\bar{V}_k'$ ,$\bar{V}_k''$ and $\bar{V}_k^{(n)}$ denote the first, second and $n$-th derivatives of $\bar{V}_k(r)$, respectively.

We may use the expansion (\ref{Kseries}) for the function $K$ in (\ref{GfromK}) and carry out the $p$-integration. Since
\begin{eqnarray}
\int _{-\infty }^{\infty }\frac{dp}{2 \pi } K_0\!\left(r,q=\frac{p}{l},t=sl^2\right)=\frac{1}{\sqrt{4 \pi s}} e^{-s l^2 \frac{4}{r^2}},
\end{eqnarray}
the result is the following $\frac{1}{l}$-expansion of the quantity $G_{l,j}(r,r;s)$ (which can be used even when $sl^2\sim O(1)$):
\begin{eqnarray}
&& G_{l,j}(r,r;s) = \frac{e^{-s l^2 \frac{4}{r^2}}}{\sqrt{4 \pi s}} \left\{ 1+\frac{1}{l} \left[ -sl^2 \bar{V}_1\right] + \frac{1}{l^2} \left[\frac{(sl^2)^2}{2} \bar{V}_1^2-s l^2 \bar{V}_2+\frac{16 (sl^2)^3}{3 r^6}-\frac{4 (sl^2)^2}{r^4} \right] \right. \nonumber\\[6pt]
&& \quad +\frac{1}{l^3} \left[ -\frac{4 (sl^2)^3}{3 r^3} \bar{V}_1'+\left(\frac{4 (sl^2)^3}{r^4}-\frac{16 (sl^2)^4}{3 r^6}\right)\bar{V}_1-\frac{(sl^2)^3}{6} \bar{V}_1^3+\frac{(sl^2)^2}{2} \left( \bar{V}_1 \bar{V}_2+ \bar{V}_2 \bar{V}_1 \right) \right. \nonumber\\[6pt]
&& \qquad \left. -\frac{(sl^2)^2}{6} \bar{V}_1'' -sl^2 \bar{V}_3 \right] + \frac{1}{l^4} \left[ \frac{2 (sl^2)^4}{3 r^3} \left( \bar{V}_1 \bar{V}_1'+\bar{V}_1' \bar{V}_1\right)+\left(\frac{8 (sl^2)^5}{3 r^6}-\frac{2 (sl^2)^4}{r^4}\right)\bar{V}_1^2 \right. \nonumber\\[6pt]
&& \qquad -\frac{4 (sl^2)^3}{3 r^3} \bar{V}_2'+\frac{(sl^2)^3}{6} \left( \frac{\left(\bar{V}_1'\right)^2}{2}+\frac{\bar{V}_1 \bar{V}_1''}{2}+\frac{\bar{V}_1'' \bar{V}_1}{2}-\bar{V}_1^2 \bar{V}_2-\bar{V}_2 \bar{V}_1^2-\bar{V}_1 \bar{V}_2 \bar{V}_1\right)+\frac{(sl^2)^4}{24} \bar{V}_1^4 \nonumber\\[6pt]
&& \qquad +\left(\frac{4 (sl^2)^3}{r^4}-\frac{16 (sl^2)^4}{3 r^6}\right)\bar{V}_2+\frac{(sl^2)^2}{2} \left( \bar{V}_2^2+\bar{V}_1 \bar{V}_3+\bar{V}_3 \bar{V}_1-\frac{\bar{V}_2''}{3} \right)-sl^2 \bar{V}_4+\frac{128 (sl^2)^6}{9 r^{12}} \nonumber\\[6pt]
&& \qquad \left.\left. -\frac{704 (sl^2)^5}{15 r^{10}}+\frac{40 (sl^2)^4}{r^8}-\frac{8 (sl^2)^3}{r^6} \right] +O\!\left(\frac{1}{l^5}\right) \right\}. \label{WKBresult}
\end{eqnarray}
The (matrix) functions $\bar{V}_1(r),\bar{V}_2(r),\cdots$ here should be found through (\ref{Vseries}), for the potential $\mathcal{V}_{l,j}$ given by (\ref{Vform1})-(\ref{Vform2}); hence, $\bar{V}_1(r),\bar{V}_2(r),\cdots$ can be expressed in terms of our profile function $f(r)$ (see(\ref{Amuform})).

Using the form (\ref{WKBresult}) with (\ref{calGdef}), the systematic large-$l$ approximation of the quantity $\mathcal{G}_l(r,r;s)$ can also be obtained. We may then use the result together with (\ref{GammaH}) to evaluate $\Gamma_{l>L}^{(-)}(A;m)$, according to the general procedure we detailed already in \cite{rea1,rea2,rea3}. That is, the $l$-sum appearing can be performed with the help of the Euler-Maclaurin summation formula and this is followed by the integration over the proper-time variable $s$. One can then verify that the $\Lambda\to\infty$ limit is indeed well-defined, for the potentially-divergent proper-time integral gets canceled by the renormalization counterterm contribution. After these, somewhat lengthy but straightforward, calculations, we have found that the quantity $\Gamma_{l>L}^{(-)}(A;m)$ in the gauge background (\ref{Amuform}) can be expressed as a $\frac{1}{L}$-series of the form
\begin{eqnarray}
&& \Gamma_{l>L}^{(-)}(A;m) = \int _0^{\infty }dr \left[Q_2(r) L^2+Q_1(r) L \right. \nonumber\\
&& \qquad \left. +Q_{\log }(r) \ln\!\left(\frac{2 L (u+1)}{\mu r}\right)+Q_0(r)+O\!\left( \frac{1}{L} \right) \right], \label{GammaHresult}
\end{eqnarray}
where $u \equiv \sqrt{1+\left(\frac{m r}{2 L}\right)^2}$
\begin{eqnarray}
&& Q_2(r)=\frac{4 r}{u} f(r) \left(r^2 f(r)-1\right), \nonumber\\[6pt]
&& Q_1(r)=\frac{3 r}{u^3} \left(3 u^2-1\right) f(r) \left(r^2 f(r)-1\right), \nonumber\\[6pt]
&& Q_{\log }(r)=\frac{r^3}{2} \left\{f(r)^2 \left(6 r^3 f'(r)-20\right)-r^2 f'(r)^2-10 r f(r) f'(r)-4 r^4 f(r)^4+20 r^2 f(r)^3\right\}, \nonumber\\[6pt]
&& Q_0(r)=\frac{r}{24 u^7} \left[3 r^2 f(r)^2 \left(-24 r^3 u^6 f'(r)+128 u^6-55 u^4+30 u^2+5\right) \right. \label{GammaHresultexplicit}\\
&& \qquad +2 r u^2 \left\{-r u^2 f''(r)+r^3 u^2 \left(6 u^2+1\right) f'(r)^2-3 \left(2 u^4+u^2+1\right) f'(r)\right\} \nonumber\\
&& \qquad +f(r) \left\{4 r^4 u^4 f''(r)+4 r^3 \left(30 u^4+5 u^2+3\right) u^2 f'(r)-3 \left(56 u^6-57 u^4+24 u^2+5\right)\right\} \nonumber\\
&& \qquad \left. +2 r^6 u^2 \left(24 u^4-13 u^2+3\right) f(r)^4-4 r^4 u^2 \left(60 u^4-7 u^2+3\right) f(r)^3\right]. \nonumber
\end{eqnarray}
If the formula (\ref{GammaHresult}) is used to our Case I where $f(r)=\frac{1}{r^2} \frac{(r/\rho)^{2 \alpha }}{(r/\rho)^{2 \alpha }+1}\; (\alpha \geq 1)$, one obtains, after the $r$-integration, the following result:
\begin{eqnarray}
\Gamma _{l>L}^{(-)}(A;m=0)=-\frac{2 L^2}{\alpha }-\frac{3 L}{\alpha }-\ln\!
\left(\frac{4 L}{\mu \rho}\right) \left(\frac{\alpha }{6}+\frac{1}{6 \alpha }+\frac{1}{2}\right)+\frac{5 \alpha }{36}-\frac{47}{72 \alpha }+O\!\left(\frac{1}{L}\right). \label{GammaHresult1}
\end{eqnarray}

For the mathematical validity of our formula (\ref{devideGamma}), we do not need to know, in the right hand side of (\ref{GammaHresult}), the explicit form of the $O(\frac{1}{L})$ term or beyond, i.e., $Q_{-1}(r) \frac{1}{L} + Q_{-2}(r) \frac{1}{L^2} + \cdots$. But, as noted in \cite{rea1,rea2,rea3}, these $\frac{1}{L}$-suppressed terms can be important to \emph{accelerate the convergence} of our calculational scheme if the low partial wave contribution has to be evaluated by numerical methods. They can of course be found by keeping further higher-order terms in (\ref{Kseries}) and (\ref{WKBresult}) above. The end results turned out to be quite lengthy: see Appendix \ref{appendixA} for the explicit expressions of $Q_{-1}(r)$ and $Q_{-2}(r)$. (These results are utilized in Sec. \ref{sec4}).

\section{Low partial-wave contribution and the full effective action} \label{sec4}
Our next task is to evaluate the low partial-wave contribution $\Gamma_{l\leq L}^{(-)}(A;m)$, given in (\ref{GammaL}), for the background field of the form (\ref{Amuform2}) with certain specific $f(r)$. By combining this with the high partial-wave contribution (calculated already in Sec. \ref{sec3}) \emph{a la} (\ref{devideGamma}), we can determine $\Gamma_{\text{ren}}^{(-)}(A;m)$; then, by (\ref{chiralformula}), the full fermion effective action $\Gamma_{\text{ren}}(A;m)$ follows at once. We use the Gel'fand-Yaglom (GY) method \cite{gypaper} to determine the ratio of two functional determinants in (\ref{GammaL}). Applying this method to functional determinants involving ordinary, non-matrix-type, differential operators is well-known, and we can thus write \cite{rea2}
\begin{eqnarray}
\frac{\det(\mathcal{H}_{l,j}+m^2)}{\det(-\partial_{(l)}^2+m^2)} = \lim_{R_e\to\infty} \frac{\psi_{l,j}(R_e)}{\psi_l^{\text{free}}(R_e)}, \qquad (\text{if } j\neq l, \text{ or } l=j=0), \label{gyformula}
\end{eqnarray}
where $\psi_{l,j}(r)$ and $\psi_l^{\text{free}}(r)$ denote the solutions to the radial differential equations
\begin{eqnarray}
&& (\mathcal{H}_{l,j}+m^2 ) \psi_{l,j}(r) = 0, \label{gyequation}\\
&& (-\partial_{(l)}^2+m^2 ) \psi_l^{\text{free}}(r) = 0 \nonumber
\end{eqnarray}
with following small-$r$ limit behaviors
\begin{eqnarray}
r\to 0 \;:\; \psi_{l,j}(r) \sim 1 \cdot r^{2l}, \qquad \psi_l^{\text{free}}(r) \sim 1 \cdot r^{2l}.
\end{eqnarray}
But, in (\ref{GammaL}), there are also functional determinants involving $2\times2$ matrix differential operators, i.e., $\mbox{\boldmath$\mathcal{H}$}_{l,l}+m^2$. For these functional determinants, we have to use the generalized GY formula \cite{kirsten}
\begin{eqnarray}
\frac{\det(\mbox{\boldmath$\mathcal{H}$}_{l,l}+m^2)}{\left\{\det(-\partial_{(l)}^2+m^2)\right\}^2} = \lim_{R_e\to\infty} \frac{\det(\psi_{l,\alpha\beta}(R_e))}{\psi_l^{\text{free}}(R_e)^2}, \label{gyformula2}
\end{eqnarray}
where $\psi_{l,\alpha\beta}(r)$ ($\alpha,\beta=1,2$) denote the solutions to the differential equations
\begin{eqnarray}
&& (\mbox{\boldmath$\mathcal{H}$}_{l,l}+m^2 )_{\alpha\beta} \psi_{l,\beta 1}(r) = 0, \nonumber\\
&& (\mbox{\boldmath$\mathcal{H}$}_{l,l}+m^2 )_{\alpha\beta} \psi_{l,\beta 2}(r) = 0 \label{coupledeq}
\end{eqnarray}
with following small-$r$ limit behaviors
\begin{eqnarray}
r\to 0 \;:\; \left(
\begin{array}{cc}
 \psi _{l,11}(r) & \psi _{l,12}(r) \\
 \psi _{l,21}(r) & \psi _{l,22}(r)
\end{array}
\right)\sim \left(
\begin{array}{cc}
 1 & 0 \\[-6pt]
 0 & 1
\end{array}
\right) r^{2 l}. \label{matrixBC}
\end{eqnarray}
Note that the two equations in (\ref{coupledeq}) may be written as a single matrix differential equation
\begin{eqnarray}
\left( \mbox{\boldmath$\mathcal{H}$}_{l,l} + m^2 \right) \mbox{\boldmath$\Psi$}_l(r) =0, \qquad \left[ \mbox{\boldmath$\Psi$}_l (r) \equiv \left(
\begin{array}{cc}
 \psi_{l,11}(r) & \psi_{l,12}(r) \\
 \psi_{l,21}(r) & \psi_{l,22}(r)
\end{array}
\right) \right]. \label{matrixDE}
\end{eqnarray}

With nonzero mass $m$ and a generic radial function $f(r)$, exact forms of the GY wave functions $\psi_{l,j}(r)$ and $\mbox{\boldmath$\Psi$}_l(r)$ are usually not available. But, in the massless limit, a certain analytic procedure can be developed to find these wave functions. In fact, with the background field of our Case I, complete GY wave functions may be obtained by this procedure. So, below, we shall show how our general procedure can be applied to calculate the effective action in our chosen backgrounds in the massless limit first (Part A). This will then be followed by the corresponding discussion with $m\neq 0$ (Part B), which requires extensive numerical analysis.

We note that, in performing this effective action calculation, it is convenient to set the length parameter $\rho$ (entering our background fields), and sometimes the normalization scale $\mu$ also, to be equal to 1. This does not amount to a loss of generality. It is related to the fact that, from simple dimensional argument and the way the normalization scale $\mu$ enters $\Gamma_{\text{ren}}(A;m)$ (see (\ref{reneffectiveaction})), the modified effective action $\tilde{\Gamma}(A;m)$ in our backgrounds, defined by the relation
\begin{eqnarray}
\Gamma_{\text{ren}}(A;m) = \frac{2}{3} \ln (\mu\rho) \int \frac{d^4x}{(4\pi)^2}\; \mathrm{tr} (F_{\mu\nu}F_{\mu\nu}) + \tilde{\Gamma}(A;m), \label{tildeGammadef}
\end{eqnarray}
should be a function of dimensionless parameters not involving $\mu$, i.e., a function of $m\rho$ and $\alpha$ for Case I, and a function of $m\rho$, $R/\rho$ and $\beta$ for Case II. According to (\ref{tildeGammadef}), we now have that
\begin{eqnarray}
\Gamma_{\text{ren}}(A;m) |_{\rho=\mu=1} = \tilde{\Gamma}(A;m) |_{\rho=1},
\end{eqnarray}
viz. by calculating the effective action with $\rho=\mu=1$, we have calculated $\tilde{\Gamma}(A;m) |_{\rho=1}$. But the modified effective action $\tilde{\Gamma}(A;m)$ for arbitrary $\rho$-value follows from $\tilde{\Gamma}(A;m) |_{\rho=1}$ by dimensional considerations --- just regard the numbers assumed for $m$ and $R$ in $\tilde{\Gamma}(A;m) |_{\rho=1}$ as denoting the values of $m\rho$ and $R/\rho$, respectively. With the quantity $\tilde{\Gamma}(A;m)$ thus found, the corresponding effective action $\Gamma_{\text{ren}}(A;m)$ for arbitrary values of $\rho$ and $\mu$ is provided through (\ref{tildeGammadef}).

\subsection{Fermion effective action in the massless limit}\label{masslesslimit}
For $m=0$ the above GY equations exhibit a special feature of factorizability. To show this, we note that the differential operator representing $-D^2-\frac{1}{2} \eta^{(+)}_{\mu\nu a} \sigma_a F_{\mu\nu}$ in the radially symmetric background (\ref{Amuform2}) (see (\ref{radialform1})) can in fact be decomposed as the product of two linear differential operators, according to following, directly verifiable, relation
\begin{eqnarray}
&& -\frac{\partial ^2}{\partial r^2}-\frac{3}{r} \frac{\partial }{\partial r}+\frac{4}{r^2} \vec{L}^2+3 r^2 f(r)^2+8 f(r) \vec{T} \cdot \vec{L}^{(+)}+4 \left[4 f(r)+r f'(r)-2 r^2 f(r)^2\right]  \vec{S}\cdot \vec{T} \nonumber\\[6pt]
&& \qquad = -\left(\frac{\partial }{\partial r}+\frac{3}{r}+\frac{4}{r}\vec{L}^{(+)}\cdot \vec{S}+4 r f(r)\vec{S}\cdot \vec{T}\right) \left(\frac{\partial }{\partial r}-\frac{4}{r} \vec{L}^{(+)}\cdot \vec{S}-4 r f(r) \vec{S}\cdot \vec{T}\right). \label{factorized}
\end{eqnarray}
Our operator $\mathcal{H}_{l,j}$ is nothing but the restriction of this operator to the partial waves with quantum numbers $l$ and $j$. For given values of $l$ and $j$, $\vec{S}\cdot \vec{T}$ is represented by the form in (\ref{STform}); similarly, for $\vec{L}^{(+)}\cdot \vec{S}$ ($=\frac{1}{2}(\vec{J}^2-\vec{Q}^2-\vec{S}^2-2 \vec{S}\cdot \vec{T})$), we have the representation
\begin{eqnarray}
\begin{array}{lcl}
\vec{L}^{(+)}\cdot \vec{S} \longrightarrow \displaystyle{\frac{l}{2}} &,& \text{if } j=l+1, \\[6pt]
\vec{L}^{(+)}\cdot \vec{S} \longrightarrow \displaystyle{-\frac{l+1}{2}} &,& \text{if } j=l-1, \\[6pt]
\vec{L}^{(+)}\cdot \vec{S} \longrightarrow \displaystyle{-\frac{1}{2(2l+1)}} \left(
\begin{array}{cc}
 l (2 l+3) & 2\sqrt{l (l+1)} \\
 2\sqrt{l (l+1)} & (l+1)(-2l+1)
\end{array}
\right) &,& \text{if } j=l\neq 0, \\[6pt]
\vec{L}^{(+)}\cdot \vec{S} \longrightarrow 0 &,& \text{if } j=l=0.
\end{array} \label{LSform}
\end{eqnarray}
Also a remark as regards the $j=l=0$ partial wave: for our backgrounds having a nonzero Pontryagin index, there will be a normalizable zero mode of the operator (\ref{factorized}).

Based on the above observation, we can recast the massless GY equation $\mathcal{H}_{l,j} \psi(r)=0$ appropriate to $j=l+1$ (and hence $\vec{S}\cdot \vec{T} \to \frac{1}{4}$ and $\vec{L}^{(+)}\cdot \vec{S} \to \frac{l}{2}$) as
\begin{eqnarray}
\left(\frac{\partial }{\partial r}+\frac{3}{r}+\frac{2l}{r}+r f(r)\right) \left(\frac{\partial }{\partial r}-\frac{2l}{r}-r f(r)\right) \psi(r) =0.
\end{eqnarray}
Therefore, for the GY wave function, we may well look for the solution to the first-order equation
\begin{eqnarray}
\left(\frac{\partial }{\partial r}-\frac{2l}{r}-r f(r)\right) \psi(r) =0.
\end{eqnarray}
This way, the GY wave function with the correct small-$r$ behavior is obtained:
\begin{eqnarray}
\psi_{l,j=l+1}(r) = r^{2l} e^{\int_0^r r_1 f(r_1)dr_1}. \label{psiplus}
\end{eqnarray}
As for the GY equation with $j=l-1$ (and so $\vec{S}\cdot \vec{T} \to \frac{1}{4}$, $\vec{L}^{(+)}\cdot \vec{S} \to -\frac{l+1}{2}$), i.e., for the equation
\begin{eqnarray}
\left(\frac{\partial }{\partial r}+\frac{3}{r}-\frac{2(l+1)}{r}+r f(r)\right) \left(\frac{\partial }{\partial r}+\frac{2(l+1)}{r}-r f(r)\right)  \psi(r) =0, \label{lm1eq}
\end{eqnarray}
the situation is not quite the same. In the latter case, solving the first order equation
\begin{eqnarray}
\left(\frac{\partial }{\partial r}+\frac{2(l+1)}{r}-r f(r)\right) \psi_1(r) =0 \label{lm1eq2}
\end{eqnarray}
results in a solution of the form
\begin{eqnarray}
\psi_1(r) = r^{-2(l+1)} e^{\int_0^r r_1 f(r_1)dr_1},
\end{eqnarray}
which is \emph{singular} as $r\to 0$. For GY wave function we thus have to look for another kind of solution to the second-order equation (\ref{lm1eq}). For such solution $\psi_2(r)$, we here put $\psi_2(r)=\psi_1(r) a(r)$ and then use the equation (\ref{lm1eq}) (together with (\ref{lm1eq2}) for $\psi_1(r)$) to obtain
\begin{eqnarray}
\left(\frac{\partial }{\partial r}+\frac{3}{r}-\frac{2(l+1)}{r}+r f(r)\right) \psi_1(r) a'(r)=0.
\end{eqnarray}
By solving this equation we are led to the expression
\begin{eqnarray}
a'(r) = r^{4l+1} e^{-2\int_0^r r_1f(r_1)dr_1}.
\end{eqnarray}
and this in turn lead to the following form for the second solution $\psi_2(r)=\psi_1(r) a(r)$:
\begin{eqnarray}
\psi_2(r) = r^{-2(l+1)} e^{\int_0^r r_1f(r_1)dr_1} \int_0^r r_2^{4l+1} e^{-2\int_0^{r_2} r_1f(r_1)dr_1} dr_2.
\end{eqnarray}
For small $r$, this function behaves as $\psi_2(r)\sim \frac{r^{2l}}{2(2l+1)}$. Hence we can identify the appropriate GY wave function with $2(2l+1)$ times this function, i.e.,
\begin{eqnarray}
\psi_{l,j=l-1}(r) = 2(2l+1) r^{-2(l+1)} e^{\int_0^r r_1f(r_1)dr_1} \int_0^r r_2^{4l+1} e^{-2\int_0^{r_2} r_1f(r_1)dr_1} dr_2. \label{psiminus}
\end{eqnarray}

Massless functional determinants for partial waves corresponding to $j=l\pm 1$ can be evaluated using the GY wave functions in (\ref{psiplus}) and (\ref{psiminus}). Especially, for our Case I, i.e., $f(r)=\frac{1}{r^2} H(r)$ with the function $H(r)$ as given in (\ref{chosenH1}) (while taking $\rho=1$), the exact wave functions are given in terms of hypergeometric functions:
\begin{eqnarray}
\psi _{l,j=l+1}(r)&=&r^{2 l} \left(r^{2 \alpha }+1\right)^{\frac{1}{2 \alpha }}, \label{masslesspsip}\\
\psi _{l,j=l-1}(r)&=&r^{2 l} \left(r^{2 \alpha }+1\right)^{\frac{1}{2 \alpha }} {}_2F_1\!\left(\frac{1}{\alpha },\frac{2 l+1}{\alpha };\frac{2 l+1}{\alpha }+1;-r^{2 \alpha }\right). \label{masslesspsim}
\end{eqnarray}
Since we have $\psi_l^{\text{free}}(r) = r^{2l}$ with $m=0$, we may now readily calculate the asymptotic wave function ratios in (\ref{gyformula}) to conclude that
\begin{eqnarray}
\ln \!\left[ \frac{\det \mathcal{H}_{l,j=l+1}}{\det (-\partial_{(l)}^2 )} \right] \sim \ln \!\left[ \frac{\psi _{l,j=l+1}(R_e)}{\psi _l^{\text{free}}(R_e)} \right] &\sim& \ln R_e, \label{asympsip}\\[6pt]
\ln \!\left[ \frac{\det \mathcal{H}_{l,j=l-1}}{\det (-\partial_{(l)}^2 )} \right] \sim \ln \!\left[ \frac{\psi _{l,j=l-1}(R_e)}{\psi _l^{\text{free}}(R_e)} \right] &\sim& -\ln R_e + \ln \!\left( \frac{2l+1}{2l} \right), \label{asympsim}
\end{eqnarray}
viz., the corresponding functional determinants individually are not well-defined. This problem, noticed in a similar context also in Ref. \cite{insdet}, occurred because of our setting $m$ to be exactly zero. As a matter of fact, the asymptotic ratios found with $m$ set to zero are in general
 {\em not} the same as the massless limits of the asymptotic ratios calculated assuming nonzero mass. (We elaborate on this aspect in Appendix \ref{appendixB}). But, if one makes a `good' grouping of different partial-wave contributions, the two results for the group coincide \cite{insdet}. In our case, such good grouping is provided by the way we combined various partial-wave contributions in (\ref{GammaL}). (This is justified in Appendix \ref{appendixB}). With this understanding we may apply our results (\ref{asympsip}) and (\ref{asympsim}) to the particular combinations entering (\ref{GammaL}), to write
\begin{eqnarray}
\ln \!\left[ \frac{\det \mathcal{H}_{l-\frac{1}{2},j=l+\frac{1}{2}}}{\det \!\left(-\partial_{(l-\frac{1}{2})}^2 \right)} \right] + \ln \!\left[ \frac{\det \mathcal{H}_{l+\frac{1}{2},j=l-\frac{1}{2}}}{\det \!\left(-\partial_{(l+\frac{1}{2})}^2 \right)} \right] = \ln \!\left(\frac{2 l+2}{2 l+1}\right), \quad \left(l=\frac{1}{2},1,\cdots \right) \label{ljcombination}
\end{eqnarray}
and
\begin{eqnarray}
\ln \!\left[ \frac{\det \mathcal{H}_{l,j=l+1}}{\det \!\left(-\partial_{(l)}^2 \right)} \right] + \ln \!\left[ \frac{\det \mathcal{H}_{l+\frac{1}{2},j=l-\frac{1}{2}}}{\det \!\left(-\partial_{(l+\frac{1}{2})}^2 \right)} \right] = \ln \!\left(\frac{2 l+2}{2 l+1}\right), \quad \left(l=\frac{1}{2},1,\cdots \right). \label{ljcombination2}
\end{eqnarray}
[We here remark that, although (\ref{ljcombination2}) follows also from considering the small mass limit, (\ref{ljcombination}) does not; but, if the formula (\ref{ljcombination}) is used together with (\ref{Psi1asymp}) below, the value obtained for the total sum becomes also consistent with the massless limit of the corresponding massive expression]. Analogous considerations may be given to our Case II as well. But, to obtain the corresponding values for the quantity in the right hand side of (\ref{ljcombination}) or (\ref{ljcombination2}), numerical integration will be required.

Our next task is to study the functional determinant from the $j=l=0$ partial wave. In this case, $\vec{S}\cdot\vec{T} \to -\frac{3}{4}$ and $\vec{L}^{(+)}\cdot\vec{S} \to 0$ and so we have the GY equation
\begin{eqnarray}
\left( \frac{\partial}{\partial r} + \frac{3}{r} - 3rf(r) \right) \left( \frac{\partial}{\partial r} + 3rf(r) \right) \psi(r)=0.
\end{eqnarray}
The GY wave function may then be identified with the solution to the first order equation, i.e.,
\begin{eqnarray}
\psi_1(r)|_{j=l=0} = e^{-3\int_0^r r_1f(r_1)dr_1}. \label{Psi1sol}
\end{eqnarray}
But, if the function $H(r)=r^2f(r)$ approaches 1 as $r\to\infty$ (i.e., for Case I and also for Case II with $\beta>0$), we find that $\psi_1(r)|_{j=l=0} \sim \frac{1}{r^3}$ as $r\to\infty$. This corresponds to a normalizable zero mode mentioned earlier, and the related functional determinant vanishes. In this case our interest will naturally be in the expression when a small mass $m$ is included; in the GY approach, this requires the knowledge on the asymptotic behavior of the GY wave function satisfying the equation $(\mathcal{H}_{0,0}+m^2)\psi=0$ with small but nonzero $m$. The latter GY wave function, which we denote as $\psi_{j=l=0}(r)$, can also be constructed using the method of Ref. \cite{falsevacuum}. Based on such analysis (see Appendix \ref{appendixB}, especially (\ref{appBmassiveeq})-(\ref{appBmassivere})), we then obtain, say, for our Case I, the following result
\begin{eqnarray}
\ln \!\left[ \frac{\det(\mathcal{H}_{0,0}+m^2)}{\det(-\partial_{(l=0)}^2+m^2)} \right] \sim \ln \!\left[\frac{\psi _{j=l=0}(R_e)}{\psi _{l=0}^{\text{free}}(R_e)}\right]=\ln  m+\ln \!\left[\frac{\Gamma \!\left(1+\frac{1}{\alpha }\right) \Gamma \!\left(\frac{2}{\alpha }\right)}{2 \Gamma \!\left(\frac{3}{\alpha }\right)}\right],
\end{eqnarray}
when $m$ is small. As this corresponds to the very first term of the first group in (\ref{GammaL}), we may combine this result with the small-mass-limit form of the second term in the same group. The relevant result, for Case I, is (this follows from (\ref{appbplusre}))
\begin{eqnarray}
-\ln \!\left[\frac{\psi _{l=0,j=1}(R_e)}{\psi _{l=0}^{\text{free}}(R_e)}\right]=\ln  m - \ln 4.
\end{eqnarray}
Hence, in the small-$m$ limit, we have
\begin{eqnarray}
&& \ln \!\left[ \frac{\det(\mathcal{H}_{0,0}+m^2)}{\det(-\partial_{(l=0)}^2+m^2)} \right]-\ln \!\left[ \frac{\det(\mathcal{H}_{0,1}+m^2)}{\det(-\partial_{(l=0)}^2+m^2)} \right] \nonumber\\
&& \qquad = 2 \ln m + \ln\!\left[ \frac{\Gamma\! \left(1+\frac{1}{\alpha }\right) \Gamma \!\left(\frac{2}{\alpha }\right)}{8 \Gamma \!\left(\frac{3}{\alpha }\right)}\right], \qquad \text{(Case I)}. \label{smallmlowest}
\end{eqnarray}
[For our Case II with $\beta<0$, the GY wave function (\ref{Psi1sol}) has a nonzero asymptotic value (i.e., $\lim_{R\to\infty} \psi_1 (R) \neq 0$) and this limit value determines the $j=l=0$ massless functional determinant].

Let us now turn to the case $j=l\neq 0$, i.e., the case where $2\times2$ matrix differential equation (\ref{matrixDE}) is relevant. Here, in dealing with the boundary condition (\ref{matrixBC}), the non-diagonal matrix form given for $\vec{L}^{(+)}\cdot \vec{S}$ in (\ref{LSform}) is not very convenient. Therefore, we perform a unitary transformation, $\vec{L}^{(+)}\cdot \vec{S} \to U (\vec{L}^{(+)}\cdot \vec{S}) U^\dag$, with
\begin{eqnarray}
U=\frac{1}{2 l+1} \left(
\begin{array}{cc}
 -1 & 2 \sqrt{l (l+1)} \\
 2 \sqrt{l (l+1)} & 1
\end{array}
\right), \qquad \left( U U^\dag = I \right) \label{Udef}
\end{eqnarray}
to find the following diagonal form for $\vec{L}^{(+)}\cdot \vec{S}$:
\begin{eqnarray}
\vec{L}^{(+)}\cdot \vec{S} \longrightarrow \frac{1}{2} \left(
\begin{array}{cc}
 l & 0 \\
 0 & -l-1
\end{array}
\right). \label{LSdiagonal}
\end{eqnarray}
Under this unitary transformation, $\vec{S}\cdot \vec{T}$ (originally given by the form in (\ref{STform})) remains unchanged, i.e.,
\begin{eqnarray}
\vec{S}\cdot \vec{T} \longrightarrow\frac{1}{4(2l+1)} \left(
\begin{array}{cc}
 -2l-3 & 4\sqrt{l (l+1)} \\
 4\sqrt{l (l+1)} & -2 l+1
\end{array}
\right). \label{STdiagonal}
\end{eqnarray}
Then, based on the factorized form (\ref{factorized}), we may first consider the first-order matrix equation
\begin{eqnarray}
\left[\frac{\partial }{\partial r}-\frac{2}{r} \left(
\begin{array}{cc}
 l & 0 \\
 0 & -l-1
\end{array}
\right)- \frac{r f(r)}{2l+1}  \left(
\begin{array}{cc}
 -2l-3 & 4\sqrt{l (l+1)} \\
 4\sqrt{l (l+1)} & -2 l+1
\end{array}
\right)\right] \mbox{\boldmath$\Psi$} (r)=0, \label{firstordermatrixDE}
\end{eqnarray}
where $\mbox{\boldmath$\Psi$}(r)$ is a $2\times 2$ matrix (see (\ref{matrixDE})). For a general radial function $f(r)$ (assumed to be finite for $r\to 0$), it will be unwieldy to exhibit the solution to this matrix equation in an explicit manner. So just let a $2\times 2$ matrix function $\mbox{\boldmath$\Psi$}_1(r)$ denote the solution to (\ref{firstordermatrixDE}) which has the following small-$r$ behavior
\begin{eqnarray}
r\to 0 \;:\; \mbox{\boldmath$\Psi$}_1(r) \sim \left(
\begin{array}{cc}
 1\cdot r^{2l}\;\; & 0\cdot r^{-2l-2} \\
 0\cdot r^{2l}\;\; & 1\cdot r^{-2l-2}
\end{array}
\right). \label{smallpsi1}
\end{eqnarray}
A comment as regards (\ref{smallpsi1}) might be appropriate here. Note that, if (\ref{firstordermatrixDE}) were regarded as an equation for a `column vector', our `$2\times 2$ matrix' $\mbox{\boldmath$\Psi$}$ would comprise two column vector solutions to (\ref{firstordermatrixDE}). Then (\ref{smallpsi1}) is equivalent to the statement that we require two independent solutions to the column vector equation (\ref{firstordermatrixDE}), say $\Psi_1^{(1)}(r)$ and $\Psi_1^{(2)}(r)$, having the small-$r$ behaviors
\begin{eqnarray}
r\to 0 \;:\; \Psi_1^{(1)}(r) \sim \left(
\begin{array}{c}
 1 \\[-6pt]
 0
\end{array}
\right) r^{2l}, \qquad \Psi_1^{(2)}(r) \sim \left(
\begin{array}{c}
 0 \\[-6pt]
 1
\end{array}
\right) r^{-2l-2}.
\end{eqnarray}

Notice that the solution to the first-order equation (\ref{firstordermatrixDE}) is not the one satisfying our boundary condition (\ref{matrixBC}) (with the above unitary transformation taken into account). Then the full second-order GY equation for $j=l\neq 0$ should admit a different kind of solutions. To find such solution $\mbox{\boldmath$\Psi$}_2(r)$, we put $\mbox{\boldmath$\Psi$}_2(r)=\mbox{\boldmath$\Psi$}_1(r) \mathbf{A}(r)$ ($\mathbf{A}(r)$ is a $2\times 2$ matrix function to be determined) and use the form with the second-order equation
\begin{eqnarray}
\left(\frac{\partial }{\partial r}+\frac{3}{r}+\frac{4}{r} \vec{L}^{(+)}\cdot \vec{S}+4 r f(r) \vec{S}\cdot \vec{T}\right) \left(\frac{\partial }{\partial r}-\frac{4}{r} \vec{L}^{(+)}\cdot \vec{S}-4 r f(r) \vec{S}\cdot \vec{T}\right) \mbox{\boldmath$\Psi$} _1(r) \mathbf{A}(r)=0 \qquad
\end{eqnarray}
for $\vec{L}^{(+)}\cdot \vec{S}$ and $\vec{S}\cdot \vec{T}$ given by the matrices in (\ref{LSdiagonal}) and (\ref{STdiagonal}). This then reduces to the first-order equation for $\mbox{\boldmath$\Phi$}(r) \equiv \mbox{\boldmath$\Psi$}_1(r) \mathbf{A}'(r)$:
\begin{eqnarray}
\left[ \frac{\partial}{\partial r} +\frac{3}{r} + \frac{2}{r} \left(
\begin{array}{cc}
 l & 0 \\
 0 & -l-1
\end{array}
\right) + \frac{rf(r)}{2l+1} \left(
\begin{array}{cc}
 -2l-3 & 4\sqrt{l (l+1)} \\
 4\sqrt{l (l+1)} & -2 l+1
\end{array}
\right) \right] \mbox{\boldmath$\Phi$}(r)=0.
\end{eqnarray}
If $\mbox{\boldmath$\Phi$}_3(r)$ denotes the solution to this equation with the small-$r$ behavior
\begin{eqnarray}
r\to 0 \;:\; \mbox{\boldmath$\Phi$}_3(r) \sim \left(
\begin{array}{cc}
 1\cdot r^{-2l-3}\;\; & 0\cdot r^{2l-1} \\
 0\cdot r^{-2l-3}\;\; & 1\cdot r^{2l-1}
\end{array}
\right), \label{smallphi3}
\end{eqnarray}
the desired second solution $\mbox{\boldmath$\Psi$}_2(r)$ can be identified with
\begin{eqnarray}
\mbox{\boldmath$\Psi$}_2(r)=\mbox{\boldmath$\Psi$}_1(r) \int^r [\mbox{\boldmath$\Psi$}_1(r_1)]^{-1} \mbox{\boldmath$\Phi$}_3(r_1) dr_1. \label{Psi2}
\end{eqnarray}
Thanks to (\ref{smallpsi1}) and (\ref{smallphi3}), this second solution has the small-$r$ behavior
\begin{eqnarray}
r\to 0 \;:\; \mbox{\boldmath$\Psi$}_2(r) \sim \frac{1}{2(2l+1)} \left(
\begin{array}{cc}
 -1\cdot r^{-2l-2}\;\; & 0\cdot r^{2l} \\
 0\cdot r^{-2l-2}\;\; & 1\cdot r^{2l}
\end{array}
\right).
\end{eqnarray}
Now, for the solution satisfying the GY boundary condition (\ref{matrixBC}), we consider a linear superposition, i.e., $\mbox{\boldmath$\Psi$}_l(r)=\mbox{\boldmath$\Psi$}_1(r)\mathbf{C}_1 + \mbox{\boldmath$\Psi$}_2(r)\mathbf{C}_2$ where $\mathbf{C}_1$ and $\mathbf{C}_2$ are suitable constant $2\times 2$ matrices. By this consideration, we can identify the GY wave function for $j=l\neq 0$ with the expression
\begin{eqnarray}
\mbox{\boldmath$\Psi$} _l(r)=\mbox{\boldmath$\Psi$} _1(r) \left(
\begin{array}{cc}
 1 & 0 \\[-6pt]
 0 & 0
\end{array}
\right)+2(2 l+1) \mbox{\boldmath$\Psi$} _2(r) \left(
\begin{array}{cc}
 0 & 0 \\[-6pt]
 0 & 1
\end{array}
\right) \label{psilpre}
\end{eqnarray}
(in the basis where $\vec{L}^{(+)}\cdot \vec{S}$ and $\vec{S}\cdot \vec{T}$ are given as in (\ref{LSdiagonal}) and (\ref{STdiagonal})). Note that, if two column vectors $\Psi_1^{(1)}(r)$ and $\Psi_1^{(2)}(r)$ ($\Psi_2^{(1)}(r)$ and $\Psi_2^{(2)}(r)$) are used to represent our solution $\mbox{\boldmath$\Psi$}_1(r)$ ($\mbox{\boldmath$\Psi$}_2(r)$), (\ref{psilpre}) identifies the desired GY solution $\mbox{\boldmath$\Psi$}_l(r)$ with the matrix formed by two column vectors $\Psi_1^{(1)}(r)$ and $2(2l+1)\Psi_2^{(2)}(r)$.

For our Case I the above matrix functions $\mbox{\boldmath$\Psi$}_1(r)$ and $\mbox{\boldmath$\Phi$}_3(r)$ can be found explicitly: if $\Psi_1(r)_{nm}(\Phi_3(r)_{nm})$ denotes the $n$th column and $m$th row of $\mbox{\boldmath$\Psi$}_1(r)(\mbox{\boldmath$\Phi$}_3(r))$, we have
\begin{eqnarray}
\Psi_1(r)_{11}&=& r^{2 l} \left(r^{2 \alpha }+1\right)^{-\frac{3}{2 \alpha }} {} _2F_1\!\left(-\frac{1}{\alpha },\frac{2 l}{\alpha };\frac{2 l+1}{\alpha };-r^{2 \alpha }\right), \nonumber\\
\Psi_1(r)_{12}&=& -\frac{2 \sqrt{l (l+1)} \left(r^{2 \alpha }+1\right)^{-\frac{3}{2 \alpha }} r^{2 \alpha -2 l-2}}{(2 l+1) (-\alpha +2 l+1)} {} _2F_1\!\left(\frac{\alpha -1}{\alpha },1-\frac{2 l+2}{\alpha };2-\frac{2 l+1}{\alpha };-r^{2 \alpha }\right), \nonumber\\
\Psi_1(r)_{21}&=& \frac{l r^{2 l} \left(r^{2 \alpha }+1\right)^{-\frac{3}{2 \alpha }}}{\sqrt{l (l+1)}} \left\{\left(r^{2 \alpha }+1\right) {} _2F_1\!\left(\frac{2 l}{\alpha }+1,1-\frac{1}{\alpha };\frac{2 l+1}{\alpha }+1;-r^{2 \alpha }\right) \right. \nonumber\\
&& \qquad\qquad\qquad\qquad\left. -{} _2F_1\!\left(-\frac{1}{\alpha },\frac{2 l}{\alpha };\frac{2 l+1}{\alpha };-r^{2 \alpha }\right)\right\}, \label{Psi1explicit}\\
\Psi_1(r)_{22}&=& \frac{r^{-2 (l+1)} \left(r^{2 \alpha }+1\right)^{-\frac{3}{2 \alpha }}}{2 (2 l+1) (-\alpha +2 l+1)} \left\{\frac{2 (\alpha -1) (2 l+1) (-\alpha +2 l+2) \left(r^{2 \alpha }+1\right) r^{2 \alpha }}{-2 \alpha +2 l+1} \right. \nonumber\\
&& \times {} _2F_1\!\left(2-\frac{1}{\alpha },2-\frac{2 l+2}{\alpha };3-\frac{2 l+1}{\alpha };-r^{2 \alpha }\right)+
\Bigg(2 (2 l+1) (-\alpha +2 l+1)  \nonumber\\
&& \left. -2 [\alpha +2 l (\alpha -2 l-3)-1] r^{2 \alpha }\Bigg) {} _2F_1\!\left(1-\frac{1}{\alpha },1-\frac{2 l+2}{\alpha };2-\frac{2 l+1}{\alpha };-r^{2 \alpha }\right)\right\}, \nonumber
\end{eqnarray}
and
\begin{eqnarray}
&& \Phi_3(r)_{11}= r^{-2 l-3} \left(r^{2 \alpha }+1\right)^{\frac{3}{2 \alpha }} {} _2F_1 \!\left(\frac{1}{\alpha },-\frac{2 l}{\alpha };-\frac{2 l+1}{\alpha };-r^{2 \alpha }\right), \nonumber\\
&& \Phi_3(r)_{12}= -\frac{2 \sqrt{l (l+1)} \left(r^{2 \alpha }+1\right)^{\frac{3}{2 \alpha }} r^{2 \alpha +2 l-1}}{(2 l+1) (\alpha +2 l+1)} {} _2F_1 \!\left(1+\frac{1}{\alpha },1+\frac{2 l+2}{\alpha };2+\frac{2 l+1}{\alpha };-r^{2 \alpha }\right), \nonumber\\
&& \Phi_3(r)_{21}= \frac{l r^{-2 l-3} \left(r^{2 \alpha }+1\right)^{\frac{3}{2 \alpha }}}{\sqrt{l (l+1)}} \left\{\left(r^{2 \alpha }+1\right) {} _2F_1 \!\left(1+\frac{1}{\alpha },1-\frac{2 l}{\alpha };1-\frac{2 l+1}{\alpha };-r^{2 \alpha }\right) \right. \nonumber\\
&& \qquad\qquad\qquad\qquad\qquad\qquad\qquad\left.-{} _2F_1 \!\left(\frac{1}{\alpha },-\frac{2 l}{\alpha };-\frac{2 l+1}{\alpha };-r^{2 \alpha }\right)\right\} \label{Phi3explicit} , \\
&& \Phi_3(r)_{22}= \frac{r^{2 l-1} \left(r^{2 \alpha }+1\right)^{\frac{3}{2 \alpha }}}{2 (2 l+1) (\alpha +2 l+1)}
\Bigg\{\left(2 (2 l+1) (\alpha +2 l+1)+2 [\alpha +2 l (\alpha +2 l+3)+1] r^{2 \alpha }\right)  \nonumber\\
&& \qquad\qquad\qquad\qquad\qquad\qquad\qquad\qquad\times {} _2F_1 \!\left(1+\frac{1}{\alpha },1+\frac{2 l+2}{\alpha };2+\frac{2 l+1}{\alpha };-r^{2 \alpha }\right) \nonumber\\
&&  -\frac{2 (\alpha +1) (2 l+1) (\alpha +2 l+2) r^{2 \alpha } \left(r^{2 \alpha }+1\right)}{2 \alpha +2 l+1} {} _2F_1 \!\left(2+\frac{1}{\alpha },2+\frac{2 l+2}{\alpha };3+\frac{2 l+1}{\alpha };-r^{2 \alpha }\right)\Bigg\} . \nonumber
\end{eqnarray}
The second matrix solution $\mbox{\boldmath$\Psi$}_2(r)$, and also the GY wave function $\mbox{\boldmath$\Psi$}_l(r)$, will be given using these results. But, for the determinant ratio (\ref{gyformula2}), all we need to know is the form of $\det \mbox{\boldmath$\Psi$}_l(r)$ for large $r=R_e$, and, because of (\ref{Psi2}) and (\ref{psilpre}), it can be recast as
\begin{eqnarray}
\det \mbox{\boldmath$\Psi$} _l(R_e) =2(2 l+1) \left( \det \mbox{\boldmath$\Psi$} _1(R_e)\right) \int^{R_e} \left([\mbox{\boldmath$\Psi$} _1(r_1)]^{-1} \mbox{\boldmath$\Phi$} _3(r_1)\right)_{22} \, dr_1. \label{Psi1easy}
\end{eqnarray}
Here, $\det \mbox{\boldmath$\Psi$}_1(R)$ is easily calculated (without using the explicit forms in (\ref{Psi1explicit})) if one notes that, for $\mbox{\boldmath$\Psi$}_1(r)$ satisfying the first order equation (\ref{firstordermatrixDE}),
\begin{eqnarray}
&& \ln \!\left[ \det (r\mbox{\boldmath$\Psi$}_1(r)) \right] = \mathrm{tr} \ln \!\left( r\mbox{\boldmath$\Psi$}_1(r)\right) \nonumber\\
&& \qquad\qquad = \int_0^r \mathrm{tr}\! \left[ \frac{1}{r} + \mbox{\boldmath$\Psi$}_1'(r) \mbox{\boldmath$\Psi$}_1(r)^{-1} \right] dr \nonumber\\
&& \qquad\qquad = \int_0^r \mathrm{tr}\! \left[ \frac{1}{r}+\frac{2}{r} \left(
\begin{array}{cc}
 l & 0 \\
 0 & -l-1
\end{array}
\right)+ \frac{r f(r)}{2 l+1} \left(
\begin{array}{cc}
 -2 l-3 & 4\sqrt{l (l+1)} \\
 4\sqrt{l (l+1)} & -2 l+1
\end{array}
\right) \right] dr \nonumber\\
&& \qquad\qquad = -2 \int_0^r r f(r) dr.
\end{eqnarray}
Hence, for $f(r) = \frac{r^{2\alpha}}{r^2 (1+r^{2\alpha})}$,
\begin{eqnarray}
\ln \!\left[ \det(r\mbox{\boldmath$\Psi$}_1(r))\right] = -\frac{1}{\alpha} \ln \!\left( 1+r^{2\alpha}\right),
\end{eqnarray}
and accordingly
\begin{eqnarray}
\det \mbox{\boldmath$\Psi$}_1(R_e) \sim 1 \cdot \frac{1}{R_e^4}. \label{Psi1asymp}
\end{eqnarray}
On the other hand, from (\ref{Psi1explicit}) and (\ref{Phi3explicit}), we find
\begin{eqnarray}
r\to\infty \;:\; \left( \mbox{\boldmath$\Psi$}_1(r)^{-1} \mbox{\boldmath$\Phi$}_3(r) \right)_{22} \longrightarrow \frac{(2 l+1)^2 r^{4 l+3} \Gamma\! \left(\frac{2 l+1}{\alpha }\right)^4}{4 l (l+1) \Gamma\! \left(\frac{2 l}{\alpha }\right)^2 \Gamma \!\left(\frac{2 l+2}{\alpha }\right)^2}.
\end{eqnarray}
Using these results in (\ref{Psi1easy}) gives rise to
\begin{eqnarray}
\det \mbox{\boldmath$\Psi$}_l(R_e) \sim \frac{(2 l+1)^3 R_e^{4 l} \Gamma \!\left(\frac{2 l+1}{\alpha }\right)^4}{8 l (l+1)^2 \Gamma\! \left(\frac{2 l}{\alpha }\right)^2 \Gamma \!\left(\frac{2 l+2}{\alpha }\right)^2}. \label{detpsi1asymp}
\end{eqnarray}
Based on this, we have
\begin{eqnarray}
\ln\!\left[ \frac{\det \mbox{\boldmath$\mathcal{H}$}_{l,l}}{\left\{ \det (-\partial_{(l)}^2)\right\}^2}\right] \sim \ln \frac{\det \mbox{\boldmath$\Psi$}_l(R_e)}{\psi_l^{\text{free}}(R_e)^2} = \ln \!\left[ \frac{(2 l+1)^3 \Gamma \! \left(\frac{2 l+1}{\alpha }\right)^4}{8 l (l+1)^2 \Gamma\! \left(\frac{2 l}{\alpha }\right)^2 \Gamma\! \left(\frac{2 l+2}{\alpha }\right)^2}\right].
\end{eqnarray}
In (\ref{GammaL}), this result may be used in conjunction with that in (\ref{ljcombination}).

To obtain the quantity $\Gamma_{l\leq L}^{(-)} (A;m)$ in the small-mass limit, all that is needed now is to consider the sum of various partial-wave functional determinants discussed above. For the backgrounds corresponding to our Case I, we find from (\ref{smallmlowest}), (\ref{Psi1asymp}), (\ref{ljcombination}) and (\ref{ljcombination2}) the following result for the sum:
\begin{eqnarray}
&& \Gamma_{l\leq L}^{(-)} (A;m) = - \ln m
-\frac{1}{2} \ln \!\left[ \frac{\Gamma\! \left(1+\frac{1}{\alpha }\right) \Gamma\! \left(\frac{2}{\alpha }\right)}{8 \Gamma \!\left(\frac{3}{\alpha }\right)} \right]  \nonumber\\
&& \qquad -\frac{1}{2} \sum _{l=\frac{1}{2},1,\cdots }^L \left\{(2 l+1)^2 \ln \!\left[\frac{(2 l+1)^2 \Gamma \!\left(\frac{2 l+1}{\alpha }\right)^4}{4 l (l+1) \Gamma \!\left(\frac{2 l}{\alpha }\right)^2 \Gamma\! \left(\frac{2 l+2}{\alpha }\right)^2}\right]+\ln\! \left(\frac{2 l+1}{2 l+2}\right)\right\}. \label{GammaLresult}
\end{eqnarray}
For large enough $L$ this quantity can be computed as follows. Here notice that the quantity inside the curly brackets, waiting for the $l$-sum in (\ref{GammaLresult}), can be approximated for large $l$ by $-\frac{4 l}{\alpha }-\frac{2}{\alpha }-\frac{\alpha ^2+3 \alpha +1}{6 l \alpha }+O\!\left(\frac{1}{l^2}\right)$. Therefore, if $L$ is large enough, we obtain from (\ref{GammaLresult})
\begin{eqnarray}
\Gamma_{l\leq L}^{(-)} &=& \frac{2 L^2}{\alpha }+\frac{3 L}{\alpha }+\left[\ln (2 L)+\gamma \right] \left(\frac{\alpha }{6}+\frac{1}{6 \alpha }+\frac{1}{2}\right) \nonumber\\
&& -\ln m -\frac{1}{2} \ln \! \left[\frac{\Gamma \!\left(1+\frac{1}{\alpha }\right) \Gamma\! \left(\frac{2}{\alpha }\right)}{8 \Gamma\! \left(\frac{3}{\alpha }\right)}\right]+C(\alpha)+O\!\left(\frac{1}{L}\right), \label{GammaLresult2}
\end{eqnarray}
$C(\alpha)$ being given by
\begin{eqnarray}
C(\alpha) &=& -\frac{1}{2} \sum _{l=\frac{1}{2},1,\cdots }^{\infty } \left\{(2 l+1)^2 \ln\! \left[\frac{(2 l+1)^2 \Gamma \!\left(\frac{2 l+1}{\alpha }\right)^4}{4 l (l+1) \Gamma \!\left(\frac{2 l}{\alpha }\right)^2 \Gamma \!\left(\frac{2 l+2}{\alpha }\right)^2}\right]+\ln\! \left(\frac{2 l+1}{2 l+2}\right) \right. \nonumber\\
&& \qquad \left. +\frac{4 l}{\alpha }+\frac{2}{\alpha }+\frac{\alpha ^2+3 \alpha +1}{6 l \alpha }\right\}. \label{Cdef}
\end{eqnarray}
The constant $C(\alpha)$ may be evaluated numerically. Observe that $O(L^2)$, $O(L)$ and $O(\ln L)$ terms in (\ref{GammaLresult2}) match precisely those of the high partial-wave contribution given in (\ref{GammaHresult1}). We thus obtain the unambiguous result for their sum, i.e., for the quantity $\Gamma_{\text{ren}}^{(-)} (A;m)$ according to (\ref{devideGamma}):
\begin{eqnarray}
\Gamma_{\text{ren}}^{(-)} (A;m) &=& \left[\ln\! \left(\frac{\mu }{2}\right)+\gamma \right] \left(\frac{\alpha }{6}+\frac{1}{6 \alpha }+\frac{1}{2}\right)-\ln m \nonumber\\
&& -\frac{1}{2} \ln\! \left[\frac{\Gamma \!\left(1+\frac{1}{\alpha }\right) \Gamma \!\left(\frac{2}{\alpha }\right)}{8 \Gamma \!\left(\frac{3}{\alpha }\right)}\right]+\frac{5 \alpha }{36}-\frac{47}{72 \alpha } + C(\alpha). \label{gammarenresult}
\end{eqnarray}

Using the result (\ref{gammarenresult}) with (\ref{chiralformula}) then provides us with the exact expression for the small-mass-limit form of the renormalized fermion effective action: i.e., for our Case I backgrounds,
\begin{eqnarray}
\Gamma _{\text{ren}} &=&2 \Gamma _{\text{ren}}^{(-)}+\ln\! \left(\frac{m}{\mu }\right) \nonumber\\
&=& -\ln (m\rho)+\frac{\alpha ^2+1}{3 \alpha } \ln (\mu\rho )+\tilde{C}(\alpha), \label{case1final}
\end{eqnarray}
where we reinstated the $\rho$-dependences, and $\tilde{C}(\alpha)$ is given by
\begin{eqnarray}
\tilde{C}(\alpha) = (\gamma-\ln 2 ) \left(\frac{\alpha }{3}+\frac{1}{3 \alpha }+1\right)- \ln\! \left[\frac{\Gamma \!\left(1+\frac{1}{\alpha }\right) \Gamma \!\left(\frac{2}{\alpha }\right)}{8 \Gamma \!\left(\frac{3}{\alpha }\right)}\right]+\frac{5 \alpha }{18}-\frac{47}{36 \alpha } + 2C(\alpha). \label{tildeCdef}
\end{eqnarray}
Note that the numerical factor $\frac{\alpha ^2+1}{3 \alpha }$ multiplying $\ln(m\rho)$ in (\ref{case1final}) reflects the value of the classical Yang-Mills action, i.e., $ \int \frac{d^4x}{(4\pi)^2} \mathrm{tr} (F_{\mu\nu}F_{\mu\nu}) = \frac{\alpha ^2+1}{2 \alpha }$ for our backgrounds. For $\alpha=1$, i.e., when the background field corresponds to a single instanton \emph{solution}, we have $\tilde{C}(1) = 4 \zeta '(-1)+\frac{\ln 2}{3}+\frac{5}{36} = -0.291747$; then, from our expression (\ref{case1final}), the previous calculation of 't Hooft \cite{thooft} is recovered. Values of $\tilde{C}(\alpha)$ for different choices of $\alpha$ are given in TABLE \ref{table1}.
\begin{table}[ht]
\begin{tabular}{|c|c|c|c|c|c|c|c|}
  \hline
  $\alpha$ & 1 & 2 & 3 & 4 & 5 & 10 & 20 \\
  \hline
  $\tilde{C}(\alpha)$ & $-0.291747$ & $-0.269189$ & $-0.378112$ & $-0.590437$ & $-0.883495$ & $-3.16105$ & $-10.0277$ \\
  \hline
\end{tabular}
\caption{Values of $\tilde{C}(\alpha)$ for various $\alpha$.} \label{table1}
\end{table}
Especially, for large enough $\alpha$ (so that $H(r)$ may resemble a step function $\theta(r-\rho)$), one can derive from (\ref{Cdef}) an appropriate asymptotic formula for $C(\alpha)$; based on this, one gets the expression
\begin{eqnarray}
\Gamma _{\text{ren}} &=& -\ln (m\rho)+\frac{\alpha ^2+1}{3 \alpha } \ln (\mu\rho ) \nonumber\\
&& -\frac{\alpha  \ln\alpha}{3}+\left\{\frac{1}{9}-4 \zeta '(-1)-\frac{\ln 2}{3}\right\} \alpha +\ln \!\left(\frac{4}{3 \pi }\right)+O\!\left(\frac{\ln \alpha}{\alpha }\right),
\end{eqnarray}
which proves to be quite accurate, say, if $\alpha>8$.
Note that, for very large $\alpha$ (i.e., when $H(r)$ is almost step-like), this one-loop result can be very
large (the $-\frac{\alpha \ln \alpha}{3}$ term can exceed the classical action value); but then, higher-loop terms
can also be significant.

\begin{figure}[ht]
\subfigure[]{\includegraphics[scale=1.5]{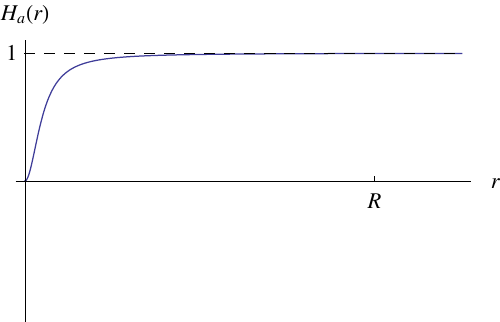}}
\subfigure[]{\includegraphics[scale=1.5]{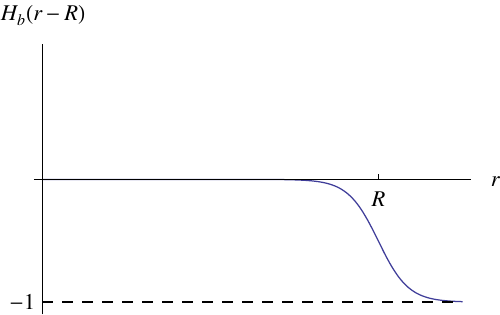}}
\caption{Schematic forms of the functions $H_a(r)$ and $H_b(r-R)$.} \label{fig3}
\end{figure}

In our Case II backgrounds the functional form of $H_{\text{II}}(r;R,\beta)$ is such that no simple expression can be obtained (even with the simplification introduced above) for the small-mass-limit form of the effective action. Hence we shall be content here with exhibiting certain feature concerning the \emph{massless} fermion effective action in our Case II backgrounds with $\beta<0$, i.e., for the case of instanton-antiinstanton-type configurations shown in Fig. \ref{fig1}(d). Actually, for the present discussion, we may take the function $H(r)\equiv r^2 f(r)$, entering the background field (\ref{Amuform2}), to have the general form
\begin{eqnarray}
H(r)=H_a(r) \left\{ 1+H_b(r-R) \right\}, \label{insainsH}
\end{eqnarray}
with the functions $H_a(r)$ and $H_b(r-R)$ broadly observing the patterns shown in Fig. \ref{fig3}.
Then it will be of interest to know the behavior of the fermion one-loop effective action as $R$, the instanton-antiinstanton separation, becomes large. With finite fermion mass, one expects that it should be approximately equal to the sum of the individual contributions from the instanton and the antiinstanton. (This is also borne out in our numerical study, presented in Part B). But, with negligible fermion mass, this is known to be generally not true \cite{callan,bardeen} --- there exists long-range interaction between the instanton and the antiinstanton.
We would like to identify such long-distance interaction term
in the massless effective action when the background field has the above form.

Since the Pontryagin index is zero for the above background, we have $\Gamma_{\text{ren}}(A;m) = 2\Gamma_{\text{ren}}^{(-)}(A;m)$. Then note that, among various partial wave contributions to $\Gamma_{\text{ren}}^{(-)}(A;m=0)$, the $l=j=0$ partial wave term is rather special. If the background field have had only the instanton part (i.e., without the $H_b$ part is (\ref{insainsH})), a normalizable zero mode would have been present in this partial wave, and hence a divergent contribution to the effective action. But with the $H_b$ part included (i.e., if an antiinstanton is also present at some distance $r=R$), there is no normalizable zero mode in any partial wave term. For the $l=j=0$ partial wave contribution, this amounts to a big change, from a divergent result to a finite one. If the instanton-antiinstanton separation $R$ becomes quite large, we must then be able to see some, strongly $R$-dependent, term (representing instanton-antiinstanton interaction at large distance) from this partial wave contribution.
Further, when the mass value is sufficiently small, our numerical study (presented in Part B of this section) shows very clearly that the contributions from other parts do not generate significant long-range interaction term.
Therefore, to extract the very long-distance interaction term, we may concentrate our study to a specific group containing the $l=j=0$ partial wave contribution of $\Gamma_{\text{ren}}^{(-)}(A;m)$, i.e., according to our grouping made in (\ref{GammaL}), to that consisting of the $l=j=0$ \emph{and} $(l=0,j=1)$ partial wave contributions.

With $m=0$ the $l=j=0$ and $(l=0,j=1)$ partial wave GY wave functions, $\psi_{0,0}(r)$ and $\psi_{0,1}(r)$, are given by (\ref{Psi1sol}) and (\ref{psiplus}):
\begin{eqnarray}
\psi_{0,0}(r) &=& e^{-3 \int_0^r r_1f(r_1) dr_1}= e^{-3 \int_0^r \frac{H(r_1)}{r_1} dr_1}, \label{l0j0sol}\\
\psi_{0,1}(r) &=& e^{\int_0^r r_1f(r_1) dr_1}= e^{\int_0^r \frac{H(r_1)}{r_1} dr_1}. \label{l0j1sol}
\end{eqnarray}
Therefore the lowest angular momentum part of our effective action expression (\ref{GammaL}) becomes
\begin{eqnarray}
&& -\frac{1}{2} \left\{ \ln \!\left[ \frac{\det \mathcal{H}_{0,0}}{\det(-\partial_{(l=0)}^2)}\right] - \ln \!\left[ \frac{\det \mathcal{H}_{0,1}}{\det(-\partial_{(l=0)}^2)}\right] \right\} = -\frac{1}{2} \ln  \!\left[ \frac{\det \mathcal{H}_{0,0}}{\det \mathcal{H}_{0,1}}\right] \nonumber\\
&& \qquad = -\frac{1}{2} \lim_{r\to\infty} \ln \!\left[\frac{\psi_{0,0}(r)}{\psi_{0,1}(r)} \right] \nonumber\\
&& \qquad = 2 \int_0^\infty \frac{H(r)}{r} dr.
\end{eqnarray}
Now, with the form (\ref{insainsH}) for $H(r)$, we may rewrite this quantity as
\begin{eqnarray}
2\int_0^\infty \frac{H(r)}{r} dr &=& 2\int_0^1 \frac{H_a(r)}{r} dr + 2\int_1^\infty \frac{H_a(r)-1}{r} dr  + 2\int_0^R \frac{H_a(r) H_b(r-R)}{r} dr \nonumber\\
&& + 2\int_R^\infty \frac{H_a(r) H_b(r-R)+1}{r} dr + 2\int_1^R \frac{1}{r} dr. \label{seperateintegral}
\end{eqnarray}
We will take $R$ to be large. Then the first two terms in the right hand side of (\ref{seperateintegral}) are finite and $R$-independent. The third and fourth terms are $R$-dependent but remains finite for large $R$. But we have also the last term, $2\int_1^R \frac{1}{r} dr = 2\ln R$, i.e., a term growing logarithmically with $R$. Based on this, we can now conclude that the massless effective action $\Gamma_{\text{ren}}(A;m)$ in the above well-separated instanton-antiinstanton background should contain a long-range logarithmic interaction term (of attractive nature), i.e.,
\begin{eqnarray}
\Gamma_{\text{ren}} \sim 4 \ln R + O(1). \label{Gammaint0leading}
\end{eqnarray}
This is consistent with the observation of Refs. \cite{callan,bardeen}.

\subsection{Fermion effective action with \boldmath$m\ne 0$}\label{numericalwork}
With $m\neq 0$, numerical integration should be considered to solve the GY equations. But, for a relatively large value of mass $m$, we have a totally different approximation scheme for the total effective action in the form of the large mass expansion. To acquire a measure on the validity of the latter scheme, we shall below summarize the appropriate formula of the large mass expansion first. Note that we here assume $\rho=\mu=1$; so without this assumption, $m$ and $R$ below become $m\rho$ and $\frac{R}{\rho}$, respectively.

\subsubsection{ Large mass expansion }\label{largemass1}
One can obtain the large mass expansion of the one-loop effective action with the help of the Schwinger-DeWitt proper time expansion. Since the related procedure is described in detail in \cite{rea1,rea2} for the case of scalar effective action, we will present only the final results that apply to our discussion with fermions. The large mass expansion for ${\Gamma}^{(-)}_{\text{ren}}(A;m)$, in the background of the form (\ref{Amuform2}), can be written as
\begin{equation}
{\Gamma}^{(-)}_{\text{ren}}(A;m)= A^{(0)}_{\rm LM} \ln m
+ A^{(2)}_{\rm LM}\frac{1}{m^2}+A^{(4)}_{\rm LM}\frac{1}{m^4}+\cdots \label{Largemass}
\end{equation}
All of the coefficients in (\ref{Largemass}) involve radial integrals of certain polynomials of the function $f(r)$ in (\ref{Amuform2}) and its derivatives. Explicitly, they are of the following forms:
\begin{eqnarray}
A^{(0)}_{\rm LM}&=&
-\frac{1}{2}\int_0^\infty dr \; r^3 \left(f^2 \left[20-6 r^3 f'\right]+r^2 f'^2+10 r f f'+4 r^4 f^4-20 r^2 f^3\right) , \\
A^{(2)}_{\rm LM}&=&\int_0^\infty dr
\left\{ -\frac{1}{40} r^4 f^{(3)} \left[-3 r f'+10 r^2 f^2-16 f\right]
-\frac{1}{120} r^2 \left[-80 r^5 f^3 f'' \right. \right. \\ \nonumber
&& \left. \left. -3 r^3 f''^2
+510 r^3 f^2 f''-432 r f f''-600 r^6 f^4 f'+60 r^5 f^2 f'^2  \right. \right.  \\ \nonumber
&&\left. \left. +30 r^4 f'^3+1120 r^4 f^3 f'+540 r^3 f f'^2+930 r^2 f^2 f'-302 r f'^2\right.\right. \\ \nonumber
&& \left.\left. -720 f f'+120 r^4 f f' f''-139 r^2 f' f''+400 r^7 f^6\right.\right. \\ \nonumber
&& \left.\left. -2160 r^5 f^5+3120 r^3 f^4-960 r f^3\right]\right\} ,  \\
A^{(4)}_{\rm LM}&=&\int_0^\infty dr
\left\{  -\frac{3}{5} r^5 f^2 f^{(4)}
-\frac{113}{40} r^4 f^2 f^{(3)}
+\frac{11}{20} r^3 f f^{(4)}+\frac{99}{56} r^2 f f^{(3)} \right.  \\  \nonumber
& & \left.  +   \frac{2}{15}   r^7 f^2 f''^2
-2 r^5 f f''^2+\frac{123}{40} r^3 f^2 f''+\frac{49}{10} r^8 f^2 f'^3
-\frac{7}{30} r^7 f'^4  \right. \\ \nonumber
&& \left. -\frac{124}{15} r^6 f f'^3-61 r^5 f^2 f'^2-\frac{173}{60} r^4 f'^3-\frac{2}{5} r^7 f^6 \left[35 r^3 f'-254\right]\right.  \\ \nonumber
&& \left. +\frac{1}{1680}r^3 \left[13 r^2 (f^{(3)})^2+1675 (f'')^2+16 r^2 f^{(4)} f''+488 r f^{(3)} f'' \right]\right. \\ \nonumber
&&\left.  +\frac{157}{840} r^4 f^{(4)} f'+\frac{299}{210} r^3 f^{(3)} f'
-\frac{1}{280} r^4 f^{(5)} \left[-2 r f'+7 r^2 f^2-11 f\right]\right. \\ \nonumber
&& \left. -\frac{13}{40} r^6 f' f''^2+\frac{41}{12} r^2 f' f''-\frac{1}{60} r^3 f f'^2 \left[36 r^4 f''-43\right]\right. \\ \nonumber
&&\left.  -\frac{2}{15} r^5 f^5 \left[25 r^4 f''-207 r^3 f'+450\right]-\frac{1}{240} r f'^2 \left[54 r^5 f^{(3)}+742 r^4 f''-175\right]\right. \\ \nonumber
&& \left. -\frac{1}{840} f f' \left[126 r^6 f^{(4)}+2226 r^5 f^{(3)}+9170 r^4 f''+495\right]\right. \\ \nonumber
&& \left. -\frac{1}{120} r^2 f^2 f' \left[8 r^5 f^{(3)}+1376 r^4 f''
-1935\right]\right.  \\ \nonumber
&&\left. +\frac{1}{15} r^4 f^3
\left[r^3 f^{(4)} -20 r^2 f^{(3)}-453 r f''+696 r^3 f'^2-1371 f'+99 r^4 f' f''\right]\right.  \\ \nonumber
&& \left. -\frac{1}{30} r^3 f^4 \left[-25 r^5 f^{(3)}-671 r^4 f''+58 r^6 f'^2-1741 r^3 f'-168\right] \right. \\ \nonumber
&& \left. +8 r^{11} f^8-52 r^9 f^7 -\frac{1}{56} r ff'' \left[14 r^5 f^{(3)}-33\right] \right\} .
\end{eqnarray}

When the radial function has the form $f(r)=r^{2\alpha-2}/(1+r^{2\alpha})$ (our Case I), it is possible to perform the radial integrals explicitly to get the result:
\begin{eqnarray}
&&  {\Gamma}^{(-)}_{\rm ren} (m) = -   \left(\frac{1}{2}+\frac{1+\alpha^2}{6|\alpha| } \right)
  \ln m    -\frac{\pi\left(24 \alpha^8-60 \alpha^6+11 \alpha^4+50 \alpha^2-25\right)}
  {1800 \alpha^6 \sin (\pi/|\alpha|)}  \frac{1}{m^2}   \nonumber  \\
&&\qquad  + \pi (\alpha^2 -4) (\alpha^2 -1) \frac{90 \alpha^8-152 \alpha^6+553 \alpha^4-126 \alpha^2-280}{11025 \alpha^8 \sin(2\pi/|\alpha|)} \frac{1}{m^4} + \cdots . \label{largemass2}
\end{eqnarray}
Note that, for $|\alpha|=1$ or $2$, taking the limit $|\alpha| \to 1$ or $|\alpha| \to 2$ in the right hand side of
(\ref{largemass2}) should be understood.
(One may compare (\ref{largemass2}) with the corresponding form for the scalar effective action given in
(3.8) of Ref. \cite{rea2}.)
This large mass expansion result  will be compared with
the numerically determined effective action later.
In Case II, it is not possible to obtain the associated radial integrals in a closed form, but  we can
evaluate them numerically.

\subsubsection{Numerically exact computation}\label{numcomput}
We now turn to our numerical evaluation method. First consider partial wave contributions with $l\neq j$ (or $l=j=0$).
By solving the differential equations (\ref{gyequation}) numerically, we can determine the value for the
ratio of two functional determinants
according to the GY formula (\ref{gyformula}).  One may easily solve the equation corresponding to the free equation.
This free radial wave function is given in terms of the modified Bessel function,
i.e., $\psi^{\rm free}_l(r)= I_{2l+1}(mr)/r$.  As noted in Ref. \cite{insdet},
it is convenient to consider the ratio of two functions
\begin{equation}
{\cal R}_{l,j}(r)=\frac{\psi_{l,j}(r) }{\psi^{\rm free}_l(r)} , \label{ratio}
\end{equation}
which has a finite value even though each of the numerator and the denominator diverges in the $r\to \infty$ limit.
This ratio function  $ {\cal R}_{l,j}(r) $
satisfies the differential equation
\begin{eqnarray}
 \frac{d^2 {\cal R}_{l,j}} {dr^2}+\left(\frac{1}{r}+2m\frac{I^\prime_{2l+1}(m r)}{I_{2l+1}(m r)}\right)
 \frac{d {\cal R}_{l,j} } {dr}  - {\cal V}_{l,j}{\cal R}_{l,j}=0,
\label{potential}
\end{eqnarray}
under the initial value boundary conditions
\begin{eqnarray}
{\cal R}_{l,j}|_{r=0}=1 , \qquad {\cal R}^\prime_{l,j}|_{r=0}=0.
\label{logbc}
\end{eqnarray}
These differential equations
share the same character as the ones one encounters in the evaluation of the
scalar effective action studied extensively  in Ref. \cite{rea2}.

In the present problem it is also necessary to evaluate functional determinants involving $2\times2$ matrix differential operators,
and for this we must solve the matrix differential equations (\ref{matrixDE})
with the boundary condition (\ref{matrixBC}). Here again,
instead of directly solving them numerically, we will consider a new matrix function
\begin{equation}
\mbox{\boldmath${\cal R}$}_{l}(r)=\frac{\mbox{\boldmath$\Psi$}_{l}(r)}{\psi^{\rm free}_l(r)}.
\end{equation}
It satisfies
the matrix differential equation of the form
\begin{eqnarray}
\frac{d^2 \mbox{\boldmath${\cal R}$}_{l}}{dr^2}+\left(\frac{1}{r}+2m\frac{I^\prime_{2l+1}(m r)}{I_{2l+1}(m r)}\right)
\frac{d \mbox{\boldmath${\cal R}$}_{l}}{dr}
-\mbox{\boldmath${\cal V}$}_{l,l}\mbox{\boldmath${\cal R}$}_{l}=0,
\end{eqnarray}
with the initial boundary conditions
\begin{equation}
\mbox{\boldmath${\cal R}$}_{l}|_{r=0}=
\left(\begin{array}{cc} 1 & 0 \\[-6pt] 0 & 1 \end{array}\right) ,
 \qquad \mbox{\boldmath${\cal R}$}_{l}^\prime|_{r=0}=\left(\begin{array}{cc} 0 & 0 \\[-6pt] 0 & 0 \end{array}\right).
\label{logbc2}
\end{equation}
Then the functional determinant of matrix differential operator in (\ref{gyformula2}) can be determined in terms of the ordinary determinant of the $2\times2$ matrix
$  \mbox{\boldmath${\cal R}$}_{l} (r= \infty)$.

Using the values $\mathcal{R}_{l,j}(r=\infty)$ and $\mbox{\boldmath$\mathcal{R}$}_l(r=\infty)$ found by the above method, each group of the functional determinants in the right hand side of (\ref{GammaL}) can be numerically evaluated to find the value of $\Gamma^{(-)}_{l\le L}(A;m)$:
\begin{eqnarray}
  \Gamma^{(-)}_{l\le L}(A;m)&=&-\frac{1}{2}\left[ \left( \ln {\cal R}_{0,0}-\ln {\cal R}_{0,1} \right)
+\sum_{l=\frac{1}{2},1,\cdots}^L \left\{ (2l+1)^2 \left( \ln \det \mbox{\boldmath${\cal R}$}_{l} \right. \right. \right. \\ \nonumber
 && \left. \left. \left. \left. +\ln  {\cal R}_{l-\frac{1}{2},l+\frac{1}{2}}
 +\ln {\cal R}_{l+\frac{1}{2},l-\frac{1}{2}}\right)   -\ln {\cal R}_{l,l+1} -\ln {\cal R}_{l+\frac{1}{2},l-\frac{1}{2}} \right\} \phantom{\frac{\frac{1}{1}}{\frac{1}{1}} }\right] \right|_{r=\infty}.
\end{eqnarray}
Combining this with the contribution from the high partial wave part (keeping up to the terms of $O(\frac{1}{L^2})$)
\begin{eqnarray}
  \Gamma^{(-)}_{l > L}(A;m)|_{\rm truncated}&=& \int_0^\infty dr \left[Q_2L^2+Q_1L+Q_{\rm log} \ln\left(\frac{2L(u+1)}{\mu r}\right)
  +Q_0 \right. \nonumber\\
 && \left. \qquad +\frac{Q_{-1}}{L}+\frac{Q_{-2}}{L^2} \right], \label{hiPwaves}
\end{eqnarray}
we can evaluate the effective action to very high accuracy. In (\ref{hiPwaves}), explicit forms of $Q_2$, \ldots  and $Q_0$ can be found from (\ref{GammaHresultexplicit}) and those of $Q_{-1}$ and  $Q_{-2}$ from Appendix \ref{appendixA}.
As explained at the end of Section \ref{sec2}, including the $\frac{1}{L}$-suppressed terms $Q_{-1}/L$ and  $Q_{-2}/L^2$ (of $\Gamma_{l>L}^{(-)}(A;m)$) in our effective action formula (\ref{devideGamma}) makes
it possible to evaluate the effective action accurately with a relatively small value of $L$.
In practice, with a choice of $20<L<50$, we could obtain the value for the effective action
with the accuracy of $10^{-6}$.

\begin{figure}
\includegraphics[scale=1.5]{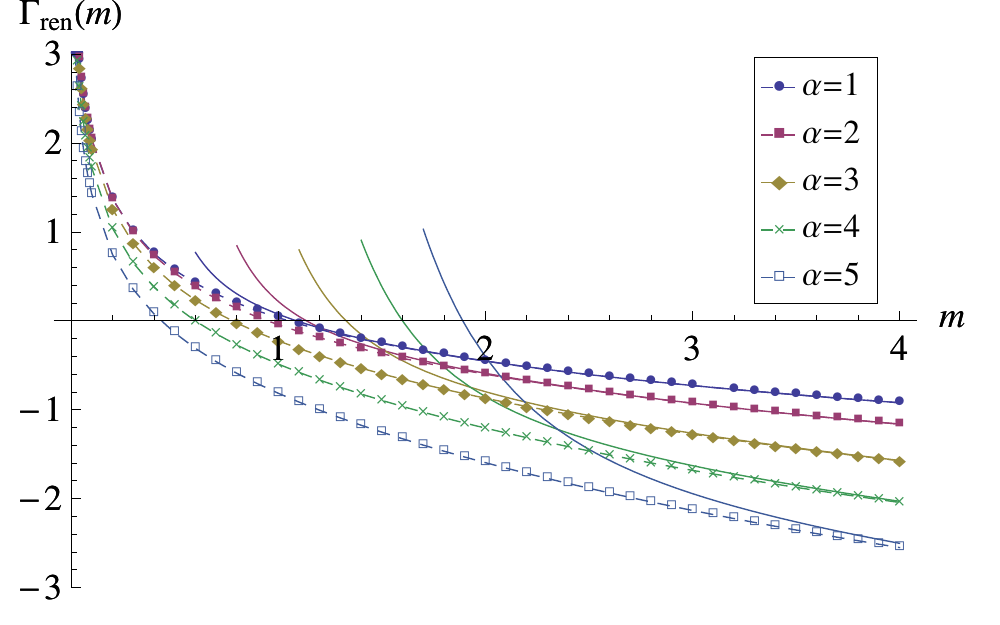}
\caption{ Plots of the  effective action in our Case I backgrounds, numerically evaluated as
a function of $m$ for various values of $\alpha$ (= 1, 2, 3, 4, 5). The solid lines (with the same color as that used for numerical values) are the corresponding results of large mass expansion. }\label{effaction}
\end{figure}

In Case I, the radial function is $f(r)=r^{2\alpha-2}/(1+r^{2\alpha})$. In this case we have the expression (\ref{GammaHresult}) for the high partial-wave part $\Gamma_{l>L}^{(-)}(A;m)$
and the $Q_{-1}$ and $Q_{-2}$ terms can  also be evaluated. This high partial-wave part must be combined with the low partial-wave part $\Gamma_{l\leq L}^{(-)}(A;m)$
which requires extensive numerical work. We have evaluated the fermion effective action as a function of $m$ for the values of $\alpha=1,2,3,4,5$ for concreteness. In Fig. \ref{effaction}, we plot these results for the full effective action $\Gamma_{\text{ren}}(A;m)$ (given by (\ref{chiralformula})), together with the corresponding results based on the large mass expansion in (\ref{Largemass}).
From this plot we see that the validity range of the large mass expansion varies with the `stiffness' $\alpha$ (see Fig. \ref{fig1}(a)) --- it is valid if the mass $m$ is such that $m\rho \gtrsim 1.2$ for $\alpha=1$ (not stiff) and $m\rho \gtrsim 4$ for $\alpha=5$ (stiff).
Note that our numerically determined curves for $\Gamma_{\text{ren}}(A;m)$ are essentially exact ones for all nonzero values of $m$.
Further, one might even try to read the values of
the quantity $\lim_{m\to 0} [\Gamma_{\text{ren}} + \ln (m\rho)]$ from the extrapolation of our numerical data. (It is quite difficult to evaluate numerically the effective action with $m=0$ since some of the solutions to the GY equations diverge in the $r\to \infty$ limit). See TABLE \ref{table2} for the values of $\tilde{C}(\alpha)$ determined numerically
for $m=\frac{1}{100}$ (assuming the `form' (\ref{case1final}) against their exact values based on (\ref{tildeCdef}).
The accuracy of the numerical method used by us is beyond question.
\begin{table}[ht]
\begin{tabular}[b]{|@{\quad}c@{\quad}|c|c|}
\hline
\multirow{2}{*}{$\alpha$} & \multicolumn{2}{c|}{$\tilde{C}(\alpha)$}\\
\cline{2-3}
& exact($m=0$) & numerical($m=\frac{1}{100}$)\\
\hline
  $1$&$-0.291747$&$-0.2916$\\
  $2$&$-0.269189$&$-0.2690$\\
  $3$&$-0.378112$&$-0.3782$\\
  $4$&$-0.590437$&$-0.5905$\\
  $5$&$-0.883495$&$-0.8835$\\
  \hline
\end{tabular}
\caption{Exact values (with $m=0$) versus numerical values (with $m=1/100$) of $\tilde{C}(\alpha)$ for $\alpha=1,2,3,4,5$} \label{table2}
\end{table}

\begin{figure}
\includegraphics[scale=1.5]{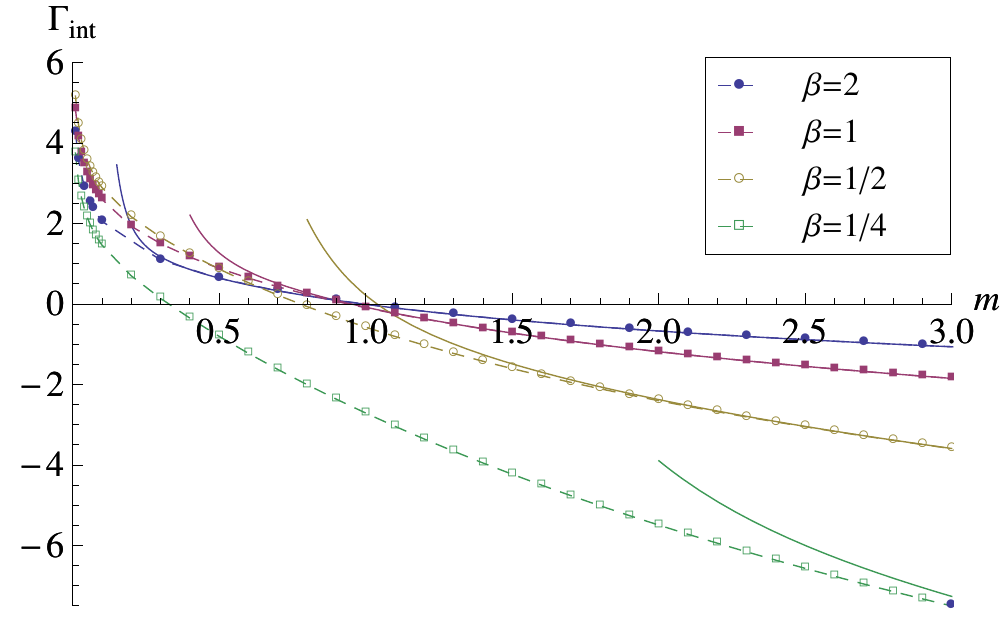}
\caption{Plots of $\Gamma_{\rm ren}(m)$ in our Case II backgrounds (with $R=5$) for  $\beta=2,\;1,\;\frac{1}{2}$ and $\frac{1}{4}$.
Near $m=0 $, all of blow up, exhibiting the $\ln (1/m)$ divergence.
The solid lines (with the same color) correspond to the results of large mass expansion.
}\label{figTp}
\end{figure}

\begin{figure}
\includegraphics[scale=1.5]{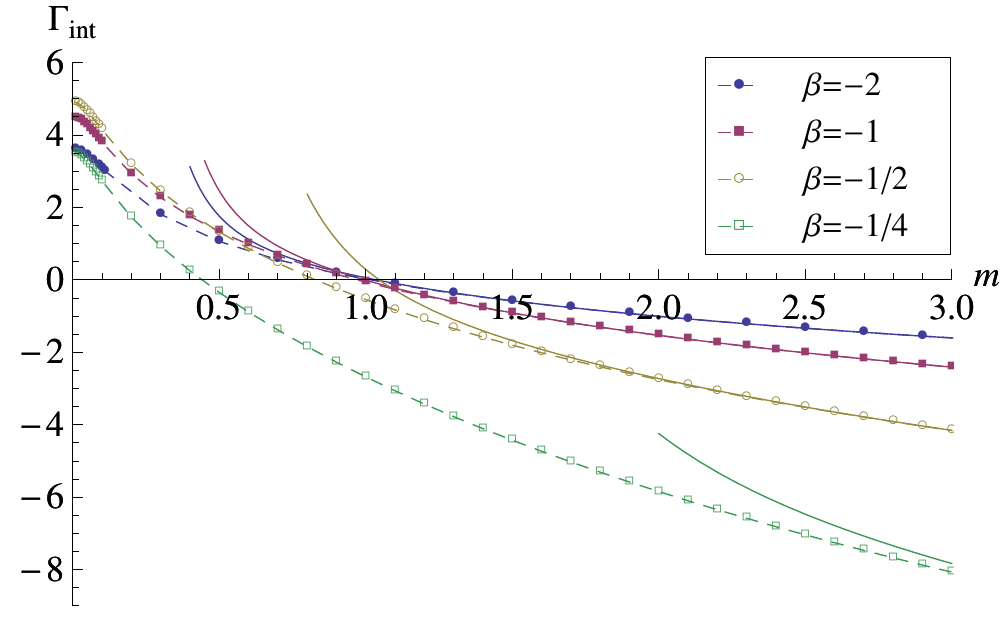}
\caption{ Plots of $\Gamma_{\rm ren}(m)$ for negative values of  $\beta$ (=$-2$,$-1$,$-1/2$, $-1/4$) when $R=5$,
together with the results of large mass expansion(represented by the solid lines). All of our 
numerical plots approach
finite values as $m\to0$.}\label{figTn}
\end{figure}

We now turn to Case II, where $f(r)=\frac{1}{2(1+r^2)}[1+\tanh(\frac{r-R}{\beta})]$. There are two free parameters $\beta$ and $R$, with an instanton-like background (see Fig. \ref{fig1}(c)) if $\beta>0$ and an instanton-antiinstanton configuration (see Fig. \ref{fig1}(d)) if $\beta<0$. For various choices of $\beta$ and $R$, we have determined numerically $\Gamma_{\text{ren}}(A;m)$, as a function of mass $m$. First, for a fixed value of $R=5$ and
$\beta=2,\;1,\;\frac{1}{2}$ and $\frac{1}{4}$, the resulting functions are plotted in Fig. \ref{figTp},
Here notice that the instanton configuration becomes more stiff-wall-like if the value of $\beta$ is reduced,
and for the validity range of large mass expansion we find $m\rho\gtrsim 1$ for $\beta=1$ and  $m\rho\gtrsim 4$ for $\beta=\frac{1}{4}$.
For the same $R$-value the fermion effective action $\Gamma_{\rm ren}$ with
negative values of $\beta(=-2,-1,-\frac{1}{2},-\frac{1}{4})$ are
evaluated also, and they are  plotted in Fig. \ref{figTn}.
Clearly, as $m$ approaches zero, the effective action becomes singular if $\beta>0$, but remains finite for $\beta<0$; this is a phenomenon directly connected with the existence or nonexistence of a fermion zero mode in the $m=0$ system.

\begin{figure}
  \includegraphics[scale=1.5]{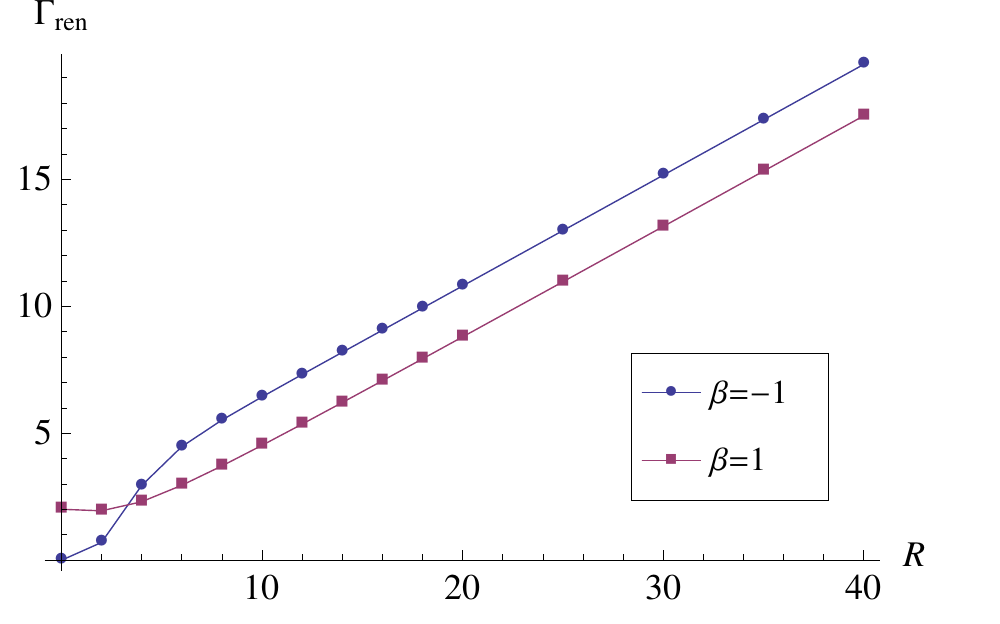}
  \caption{Plots of $\Gamma_{\rm ren}$ for the mass value $m=\frac{1}{10}$  as a function of $R$ for two cases
  with $\beta=1$ (blue dots) and $\beta=-1$ (red squares).}\label{figTR}
 \end{figure}

We also studied how the fermion effective action changes as the parameter $R$ is varied (taking here $m=\frac{1}{10}$ and $\beta=\pm1$): these results are in Fig. \ref{figTR}.
From the plots shown in Fig. \ref{figTR}, one will notice that the fermion effective actions for both values of $\beta=\pm 1$
grow linearly with $R$ if $R$ becomes large. This should not be anything surprising --- for the
gauge background field involved here, the classical Yang-Mills action, which enters the effective action through renormalization counterterms, also grows linearly with $R$. (That the two curves in Fig. \ref{figTR}, one for $\beta=1$ and the other appropriate to the case $\beta=-1$, have even the same large-$R$ \emph{slope} is related to the point discussed below).

\begin{figure}
  \includegraphics[scale=1.5]{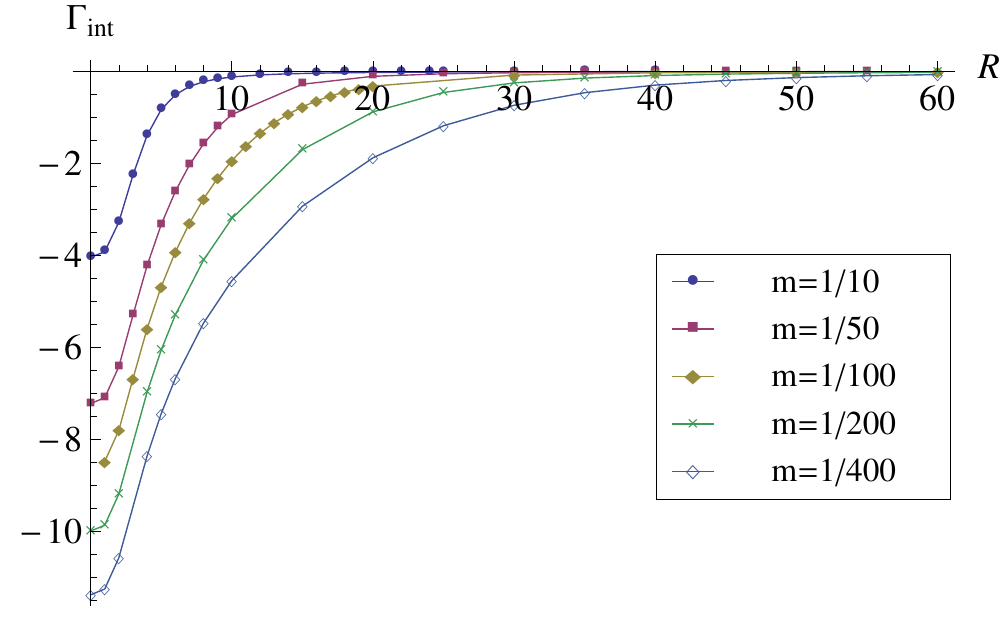}
  \caption{The interaction energy $\Gamma_{\text{int}}$ as a function the distance $R$, for $m=1/10$, 1/50, 1/100, 1/200, 1/400.}\label{figDT}
\end{figure}

Here recall that, according to the relation (\ref{HbarII}) and the remarks that follow immediately, our Case II background with a negative value of $\beta=-\beta_0$ ($\beta_0>0$) can actually be viewed as a composite configuration involving an instanton located near the origin (which is gauge-equivalent to our Case I background with $\alpha=1$) and an antiinstanton-like configuration associated to our Case II background with positive $\beta=+\beta_0$.
Then, we may define the fermion-induced `interaction energy' between the instanton and the (spherical-wall-like) antiinstanton, separated by distance $R$, as
\begin{eqnarray}
\Gamma_{\text{int}}(R;m) = \left. \Gamma_{\text{ren}}^{\text{(II)}}(R;m) \right|_{\beta=-\beta_0} - \left[ \left. \Gamma_{\text{ren}}^{\text{(I)}}(m) \right|_{\alpha=1} + \left. \Gamma_{\text{ren}}^{\text{(II)}}(R;m) \right|_{\beta=\beta_0} \right], \label{diffG}
\end{eqnarray}
where the designation (I) or (II) refer to our Case I or Case II background, respectively. Choosing $\beta_0=1$ for definiteness, we used our numerical results obtained for $\Gamma_{\text{ren}}(A;m)$ in Case II background to study how this interaction energy depends on $R$, at some chosen values of fermion mass $m$. See our plots shown in Fig. \ref{figDT}.
Clearly, with a finite value for $m$, this interaction energy vanishes as $R$ becomes sufficiently large. But this interaction dies away at far slower rate as the mass value becomes very small --- i.e., smaller the fermion mass, more long-ranged interaction seen between the instanton and antiinstanton. [As we remarked already,
with strictly zero mass, it is very difficult to perform numerical study].

\begin{figure}
  \includegraphics[scale=1.5]{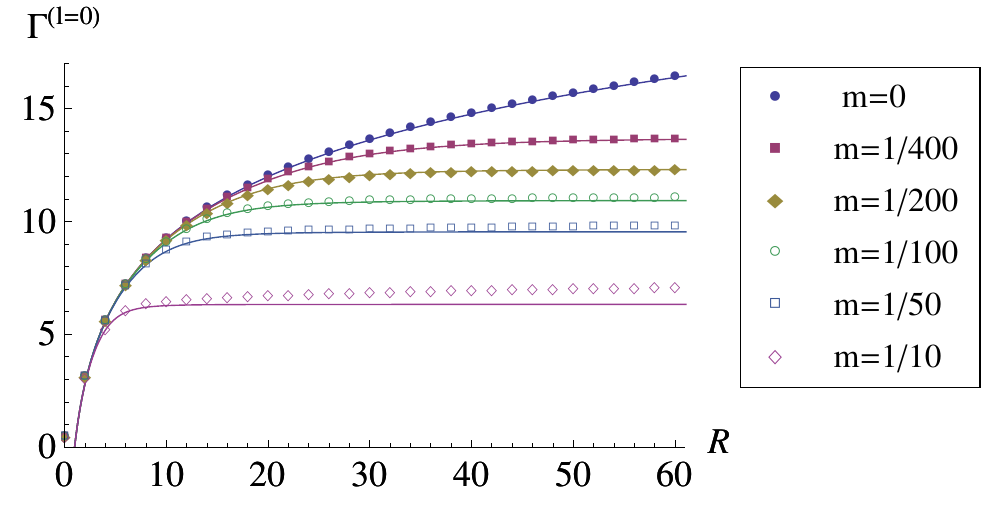}
  \caption{Contribution from the  $l=0$ partial wave term, $\Gamma^{(l=0)}(R;m)$,
   as a function of $R$ for various small values of $m$ (=$0$, $\frac{1}{400}$, $\frac{1}{200}$, $\frac{1}{100}$, $\frac{1}{50}$, $\frac{1}{10}$). The solid lines (with the appropriate color) represent our approximate formula in (\ref{approxzero}) for the given mass values.}\label{figpw0}
\end{figure}

\begin{figure}
  \includegraphics[scale=1.5]{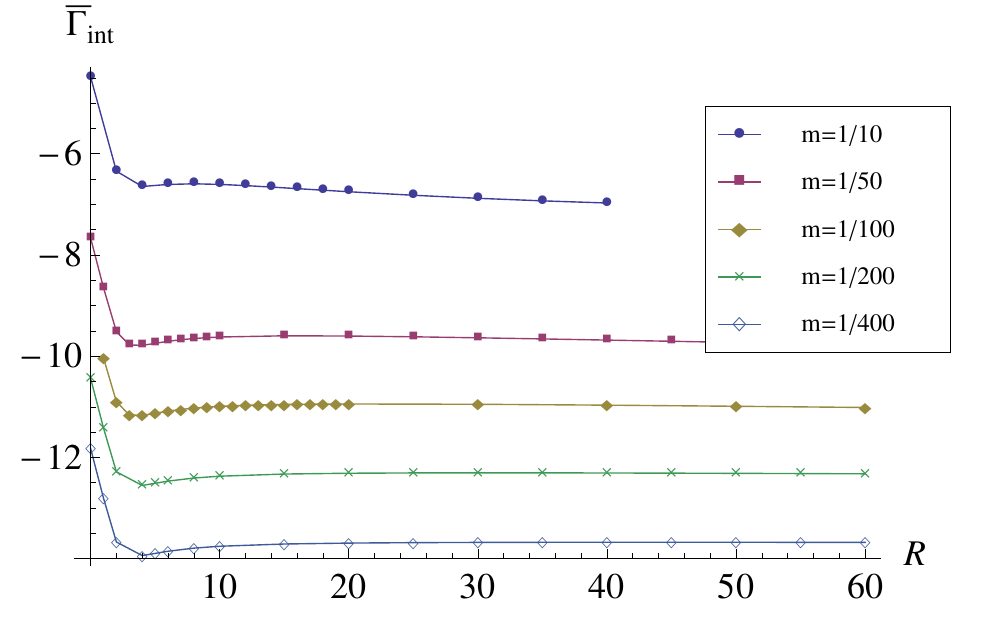}
  \caption{The interaction energy with the very $l=0$ contribution removed, $\bar{\Gamma}_{\text{int}}(R;m)$.}\label{intE2}
\end{figure}

To see the origin of the above long-range interaction at small mass, it is useful to separate the contribution to the effective action $\Gamma_{\text{ren}}^{\text{(II)}}(R;m)$ (with $\beta<0$) coming from the first brackets in the right hand side of (\ref{GammaL}) --- the part containing the $l=0$ partial waves --- from the rest. It is in this $l=0$ partial wave term (which we denote as $\Gamma^{(l=0)}(R;m)$ below) where the effects related to the disappearance of normalizable fermion zero modes (at $m=0$) in the instanton-antiinstanton composite configuration are relevant. Now see our numerical results for $\Gamma^{(l=0)}(R;m)$ (with $\beta=-1$) given in Fig. \ref{figpw0}, and see also Fig. \ref{intE2} where we present the plots for the effective action with the very $l=0$ contribution removed, i.e., for
\begin{equation}
\bar{\Gamma}_{\text{int}}(R;m) = \Gamma_{\text{int}}(R;m)- \Gamma^{(l=0)}(R;m)
\end{equation}
The fact that the latter quantity
$\bar{\Gamma}_{\text{int}}(R;m)$ as a function of $R$ becomes flat rapidly is an unambiguous sign that the $l=0$ partial wave term is mainly responsible for the above long range interaction between the instanton and the antiinstanton at small fermion mass. [The data for $m=0$ included here is the result of our direct calculation (using the exact massless GY wave functions) for the $l=0$ partial wave contribution, and at large $R$ this curve is clearly consistent with the behavior found in (\ref{Gammaint0leading})].

\begin{figure}
\includegraphics[scale=1.5]{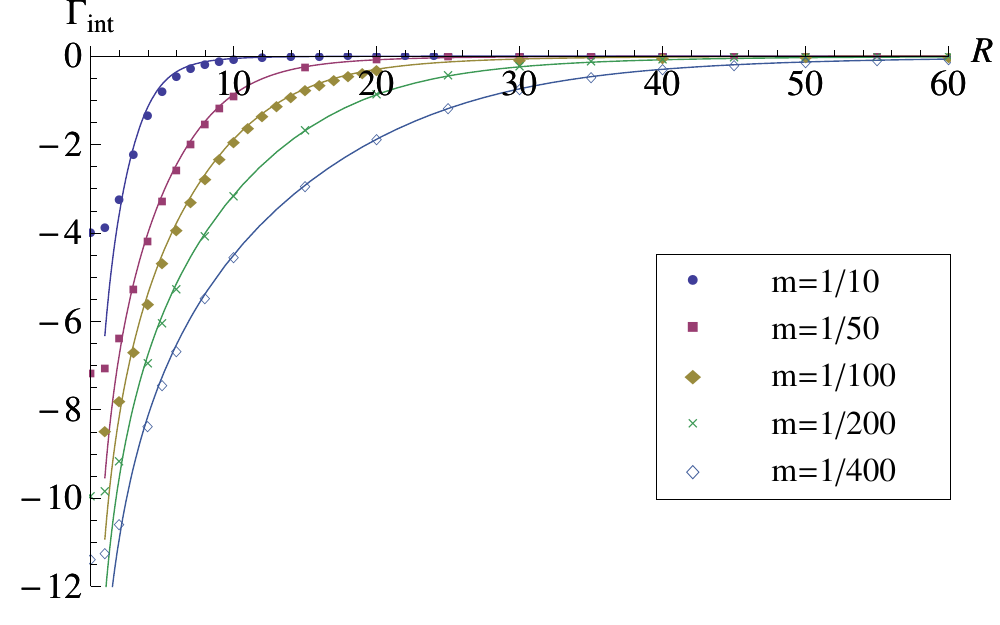}
\caption{Plot of the same data in Fig. \ref{figDT} together with our approximation function $-\ln ( 1 +A/(m^2R^4))$.}
\label{intE3}
\end{figure}

We found that the numerical data for the function $\Gamma^{(l=0)}(R;m)$ presented in Fig. \ref{figpw0} are
well approximated by the simple function
\begin{equation}
 \Gamma^{(l=0)}_{\rm approx}(R;m)= -\ln \!\left(\frac{m^2}{A} + \frac{1}{R^4}\right), \qquad (A\approx 5.55).\label{approxzero}
\end{equation}
Certainly,  it is valid when $m$ is small and $R$ is large.  As $m\to 0$
it becomes the function $4\ln R$ in (\ref{Gammaint0leading}), derived in Sec. \ref{masslesslimit}.
In the $R\to \infty$ limit, it approaches a constant, $-\ln m^2/A$. The values of $-\ln m^2/A$ for $m=1/400$, $1/200$,
$1/100$, $1/50$, and $1/10$ are 13.69, 12.31, 10.92, 9.53, and 6.31, respectively.
These values, if the opposite sign is taken, are very close to the values denoted by flat lines in Fig. \ref{intE2} (which correspond to the sum of all $l\ne0$ contributions to the interaction energy and the $l=0$ contributions to the effective actions for the separated instanton and antiinstanton configurations).
Adding $\ln m^2/A$
to the function in (\ref{approxzero}), we may thus get an approximate formula
 for the interaction energy in the form
\begin{equation}
  \Gamma_{\rm int}^{\rm (approx)}(R;m)= -\ln \!\left(1+ \frac{A}{m^2 R^4}\right). \label{intEapp}
\end{equation}
In Fig. \ref{intE3}, plots for this function for the mass values chosen for our numerical works are given together with the related numerical data for comparison.
Note that the numerical data and those our approximate formula coincide very well for $R \gtrsim 5$,
 as long as  $m \lesssim 1/10$.

\section{Concluding remarks} \label{seccon}
The partial-wave cutoff method has been extended to evaluate the 4-D spinor effective action in radially symmetric, non-Abelian, gauge backgrounds. With a suitable extension of our previous scheme, it is shown that this method retains its full power even when an infinite number of functional determinants of \emph{matrix-valued} radial differential operators have to be evaluated. By this method we have determined the fermion effective action numerically (for generic mass value) in various instanton-like and instanton-antiinstanton-like backgrounds. The validity range of large mass expansion has been checked using these numerically exact calculations. Also the results of our computation have been used to study the effective, light-fermion-induced, instanton-antiinstanton interaction. The technical aspects, elaborated in the present paper, should find useful application in other effective-action-related studies. One might also make use of our study as a basis to test how good non-Abelian derivative expansions \cite{avramidi} are. (Such study in the Abelian case was made in Ref. \cite{tba}).

In the massless limit, evaluating the fermion effective action is in some sense simpler than evaluating the scalar effective action in the same background. This is due to the unique feature of factorizability that
 the massless fermion Gel'fand-Yaglom equations for partial waves have. By utilizing this factorization fully, we have in fact found analytically the small-mass-limit form of the full fermion effective action in our general Case I backgrounds. As one can always use the systematic large mass expansion result for the effective action if mass is not so small, the additional knowledge on its small-mass-limit behavior is often sufficient to obtain a global fit to the full mass-dependence of the effective action \cite{inssmallmass}. In this regard, some further effort might be desirable in trying to solve the massless fermion Gel'fand-Yaglom equations, with more general background functions than the forms considered here.

\section*{ACKNOWLEDGMENTS}
We would like to thank G. V. Dunne for getting us interested in the instanton-antiinstanton-type configuration (our Case II background for $\beta<0$). This work was supported by the Basic Science Research Program through the National Research Foundation of Korea (NRF) funded by the Ministry of Education, Science and Technology (No. 2009-0076297 (C.L.) and No. 2010-0011223 (H.M.)).

\appendix
\section{Subleading terms for the High partial-wave contribution \boldmath$\Gamma_{l>L}^{(-)}(A;m)$} \label{appendixA}
For the background field of the form (\ref{Amuform}), (\ref{GammaHresult}) and (\ref{GammaHresultexplicit}) contain our calculated result for the high partial-wave contribution $\Gamma_{l>L}^{(-)}(A;m)$. But only the \emph{minimal} terms were included in these formulas. In case we want to implement our effective action calculation scheme numerically, it will then be necessary to evaluate a very large number of partial-wave functional determinants from the low partial-wave side. But, if we include some $\frac{1}{L}$-suppressed terms in our expression for the high partial-wave contribution, it becomes possible to take our partial wave cutoff $L$ at relatively small value; that is, based on fewer calculations of functional determinants from the low partial-wave side, we can reach numerically convergent results for the full quantity $\Gamma_{l>L}^{(-)}(A;m)$. As we utilized this idea in our numerical analysis given in Sec. \ref{sec4}, we will here present the explicit forms of the $O(\frac{1}{L})$ and $O(\frac{1}{L^2})$ terms which may be kept inside the integrand of our high partial-wave contribution formula (\ref{GammaHresult}). These terms, which we denote as $Q_{-1}(r) \frac{1}{L}+Q_{-2}(r) \frac{1}{L^2}$, are given by (here, $u=\sqrt{1+\left(\frac{m r}{2 L}\right)^2}$)
\begin{eqnarray}
&& Q_{-1}(r) = \frac{r}{32u^9} \left[ -6 r^6 u^2 \left(21 u^4-18 u^2+5\right) f^4+12 r^4 u^4 \left(u^2-1\right) f f''-6 r^2 u^4 \left(u^2-1\right) f'' \right. \nonumber\\
&& \quad +12 r^4 u^2 \left(27 u^4-12 u^2+5\right) f^3+r^2 \left(32 u^8-389 u^6+297 u^4-75 u^2-105\right) f^2 \nonumber\\
&& \quad -6 r^4 u^4 \left(u^2+1\right) \left(f'\right)^2-60 r^3 u^2 \left(u^4+1\right) f f'-2 r u^2 \left(8 u^6+3 u^4+6 u^2-15\right) f' \nonumber\\
&& \quad \left. +72 r^5 u^6 f^2 f'+\left(-64 u^8+191 u^6-225 u^4-15 u^2+105\right) f \right], \\
&& Q_{-2}(r) = \frac{r}{7680 u^{13} (u + 1)} \left[ -80 u^4 \left\{80 u^6 v-(u+1) \left(103 u^4-38 u^2+7\right)\right\} r^{10} f^6 \right. \nonumber\\
&& \quad +4800 u^8 \left(2 v u^2-u-1\right) r^9 f^4 f'-240 u^6 \left\{4 v u^4+(u+1) \left(11 u^2-5\right)\right\} r^8 f^2 \left(f'\right)^2 \nonumber\\
&& \quad +240 u^4 \left\{144 u^6 v-(u+1) \left(99 u^4-18 u^2+7\right)\right\} r^8 f^5 \nonumber\\
&& \quad +160 u^6 \left\{8 u^4 v-(u+1) \left(13 u^2-5\right)\right\} r^8 f^3 f''-480 u^{10} v r^7 \left\{ \left(f'\right)^3 + f^2 f^{(3)} \right\} \nonumber\\
&& \quad -160 u^4 \left\{112 u^6 v-5 (u-1) (u+1)^2 \left(u^2-7\right)\right\} r^7 f^3 f'-1920 u^{10} v r^7 f f' f'' \nonumber\\
&& \quad -240 u^6 \left\{36 v u^4+(u+1) \left(7 u^2+5\right)\right\} r^6 f \left(f'\right)^2+24 u^8 \left\{2 v u^2+3 (u+1)\right\} r^6 \left(f''\right)^2\nonumber\\
&& \quad -20 u^2 \left\{2496 u^{10}+(u+1) \left(3576 u^8-4377 u^6+3135 u^4-595 u^2-315\right)\right\} r^6 f^4\nonumber\\
&& \quad -240 u^6 \left\{34 u^4 v-(u+1) \left(u^2-5\right)\right\} r^6 f^2 f''+48 u^8 \left\{3 v u^2+2 (u+1)\right\} r^6 f' f^{(3)}\nonumber\\
&& \quad +48 u^8 (u+1) r^6 f f^{(4)}-240 u^4 \left\{62 u^8+(u+1) \left(8 u^6+144 u^4-5 u^2+35\right)\right\} r^5 f^2 f'\nonumber\\
&& \quad -24 u^8 (u+1) r^4 f^{(4)}+48 u^6 \left\{16 v u^4+(u+1) \left(9 u^2+10\right)\right\} r^5 f f^{(3)}\nonumber\\
&& \quad +4 u^4 \left\{1208 u^8+(u+1) \left(968 u^6+707 u^4+1190 u^2+525\right)\right\} r^4 \left(f'\right)^2\nonumber\\
&& \quad +40 u^2 \left\{384 u^{10}+(u+1) \left(1800 u^8-3561 u^6+1935 u^4-595 u^2-315\right)\right\} r^4 f^3\nonumber\\
&& \quad +8 u^4 \left\{864 u^8+(u+1) \left(1104 u^6-489 u^4+900 u^2+385\right)\right\} r^4 f f''\nonumber\\
&& \quad +16 u^6 \left\{139 v u^4+(u+1) \left(76 u^2+55\right)\right\} r^5 f' f''-40 u^6 (u+1) \left(2 u^4+3 u^2+6\right) r^3 f^{(3)}\nonumber\\
&& \quad +40 u^2 \left\{288 u^{10}+(u+1) \left(24 u^8+657 u^6-597 u^4+973 u^2+315\right)\right\} r^3 f f'\nonumber\\
&& \quad -15 (u+1) \left(3072 u^{10}-7779 u^8+3820 u^6+4270 u^4-4788 u^2-1155\right) r^2 f^2\nonumber\\
&& \quad -20 u^4 (u+1) \left(64 u^6-159 u^4+126 u^2+77\right) r^2 f''\nonumber\\
&& \quad +60 u^2 (u+1) \left(14 u^8-41 u^6+327 u^4-231 u^2-105\right) r f'\nonumber\\
&& \quad \left. +15 (u+1) \left(1040 u^{10}-2823 u^8-400 u^6+6930 u^4-3528 u^2-1155\right) f \right],
\end{eqnarray}
where $v\equiv u^2+u+1$ and $f'$, $f''$ and $f^{(n)}$ denote the first, second and $n$-th derivatives of $f(r)$.

\section{Small mass limits of Gel'fand-Yaglom wave functions} \label{appendixB}
In this appendix we will describe the method that allows to determine the asymptotics of massive GY solutions when the mass is sufficiently small. We will here assume that the exact massless GY solution is known. Then, depending on whether the massless GY solution is normalizable or not, we have two different ways to obtain the desired GY solution for small nonzero mass with global validity (i.e., the asymptotic region included).

First, suppose the exact $m=0$ solution does not correspond to a normalizable zero mode (like the GY solutions given in (\ref{masslesspsip}), (\ref{masslesspsim}) and (\ref{psilpre})). Then we can follow the idea of Ref. \cite{inssmallmass} and write the perturbative GY solution of (\ref{gyequation}) for small nonzero mass as
\begin{eqnarray}
\psi(r;m) = \psi_0(r) + m^2 \psi_1(r) + m^4 \psi_2(r) \cdots, \label{psismallm}
\end{eqnarray}
where $\psi_0(r)$ is the known massless solution, and $\psi_1(r),\psi_2(r),\cdots$ denote appropriate mass-independent functions. We will refer to this solution as the small-$r$ solution. Being concerned with the leading small mass behavior, we will here keep the leading order solution in (\ref{psismallm}) only. This naive solution works fine when the mass $m$ is very small and at the same time $mr$ can be taken to be finite. But, in our case, we are interested in the large-$r$ asymptotic behavior of $\psi(r;m)$, i.e., at $r\gg\frac{1}{m}$. Therefore we need to consider another perturbative solution which we call as the the large-$r$ solution. For this large-$r$ solution we change the variable $r$ to $x=mr$ (see Ref. \cite{inssmallmass} for more detailed discussions) and write the corresponding solution as $\varphi(x)$ (instead of $\psi(r)$). We can then recast the GY equation (\ref{gyequation}) as
\begin{eqnarray}
\left\{-\frac{\partial ^2}{\partial x^2}-\frac{3}{x} \frac{\partial }{\partial x}+\frac{4l(l+1)}{x^2}+\frac{1}{m^2} \mathcal{V}_{l,j}\!\left(\frac{x}{m}\right)+1\right\} \varphi (x)=0. \label{largereq}
\end{eqnarray}
From (\ref{Vform1})-(\ref{Vform2}) and  assuming that the function $H(r)=r^2f(r)$ approaches 1 as $r\to\infty$ (i.e., for Case I and also for Case II with $\beta>0$), we can approximate the potential in (\ref{largereq}) such that
\begin{eqnarray}
\frac{4l(l+1)}{x^2}+\frac{1}{m^2} \mathcal{V}_{l,j}\!\left(\frac{x}{m}\right) \longrightarrow \frac{4 q (q+1)}{x^2}+O\!\left(m^2\right), \label{largerpotex}
\end{eqnarray}
($q$ denotes the quantum number defined in (\ref{angnumdef})). Note that this approximation is good as long as $\frac{x}{m}$ is not so small(, say, compared to 1). With only the first term in (\ref{largerpotex}) kept, we have the leading order solution of (\ref{largereq}) in the form
\begin{eqnarray}
\varphi _0(x)=\text{(const.)} 2 (2 q+1)! \left(\frac{2}{m}\right)^{2 q} \frac{I_{2 q+1}(x)}{x}, \label{leadinglarger}
\end{eqnarray}
which is proportional to the free solution of (\ref{largereq}) with $l$ replaced by $q$. The proportionality constant will be determined at some intermediate point $r=R$ (or, equivalently, at $x=m R$), for $R$ satisfying the condition $1\ll R \ll \frac{1}{m}$. The small-$r$ solution should be valid for $0\leq r \lesssim R$, while the large-$r$ solution is valid for $r \gtrsim R$ (i.e., $x \gtrsim mR$); hence, both solutions are valid near the point $r\sim R$, so that we can demand
\begin{eqnarray}
\psi(R) = \varphi(mR). \label{demand}
\end{eqnarray}
If $\psi_0(R)$ has an asymptotic behavior $\psi_0(R) \sim c_0 R^{2q}$ (with an appropriate constant $c_0$), we can now fix the overall constant in (\ref{leadinglarger}) (i.e., demand $\varphi_0(mR)\sim c_0R^{2q}$) to obtain the complete, zeroth order, solution
\begin{eqnarray}
\varphi _0(x)=2 c_0 (2 q+1)! \left(\frac{2}{m}\right)^{2 q} \frac{I_{2 q+1}(x)}{x}. \label{leadinglargerex}
\end{eqnarray}

Based on the form (\ref{leadinglargerex}), we can obtain the correct asymptotic behavior of the GY wave function:
\begin{eqnarray}
\lim_{r\to\infty} \psi(r;m) \sim \lim_{x\to\infty} \varphi_0(x) \sim c_0 \sqrt{\frac{2}{\pi }} \left(\frac{2}{m}\right)^{2 q} (2 q+1)! \frac{e^x}{x^{3/2}}.
\end{eqnarray}
Consequently, for the ratio of GY solutions, we find
\begin{eqnarray}
\lim_{r\to\infty} \frac{\psi(r;m)}{\psi^{\text{free}}(r;m)} \sim c_0 \frac{(2 q+1)!}{(2 l+1)!} \left(\frac{2}{m}\right)^{2 q-2 l}. \label{psiasymp}
\end{eqnarray}
Note that, as $m$ goes to zero, the above ratio goes to infinity or zero or remain finite depending on the value of $q$. Let us apply this to our Case I. The functional determinants for partial waves corresponding to $j=l\pm 1$ at small mass limit can be found immediately. From (\ref{masslesspsip}) and (\ref{masslesspsim}), the large-$R$ behaviors of the massless solutions are
\begin{eqnarray}
\psi_{l,l+1}(r;m=0) &\sim& R^{2l+1}, \\
\psi_{l,l-1}(r;m=0) &\sim& \frac{2l+1}{2l} R^{2l-1}.
\end{eqnarray}
Thus, by identifying $c_0$ from these behaviors and plugging them to (\ref{psiasymp}) (with $q=l\pm\frac{1}{2}$ for $j=l\pm1$), we find that
\begin{eqnarray}
\frac{\det(\mathcal{H}_{l,l+1} +m^2)}{\det(\mathcal{H}_l^{\text{free}} +m^2)} = \lim_{r\to\infty} \frac{\psi_{l,l+1}(r;m)}{\psi_l^{\text{free}}(r;m)} &\sim& \frac{4(l+1)}{m}, \label{appbplusre}\\
\frac{\det(\mathcal{H}_{l,l-1} +m^2)}{\det(\mathcal{H}_l^{\text{free}} +m^2)} = \lim_{r\to\infty} \frac{\psi_{l,l-1}(r;m)}{\psi_l^{\text{free}}(r;m)} &\sim& \frac{m}{4l}.
\end{eqnarray}
Now one can verify explicitly that these small mass limits yield, if used for a certain specific group (indicated in the main text), the same results as those based on the exact zero mass results such as (\ref{ljcombination2}) with the same $(l,j)$ combination.

For $j=l=0$ (and when the function $H(r)=r^2f(r)$ approaches 1 as $r\to\infty$), it is rather difficult to follow the method of Ref. \cite{inssmallmass}. In this case, the massless GY solution (\ref{Psi1sol}) corresponds to the \emph{normalizable zero mode}. As a result, the small-$r$ solution cannot be matched with the large-$r$ solution of the form (\ref{leadinglarger}). Here we need to consider the Macdonald function for the large-$r$ solution; but, the Macdonald function vanishes exponentially as $r\to\infty$, and therefore we need to take into account the next order term which is mathematically very complicated. In this situation the method of Ref. \cite{falsevacuum}, which utilizes the normalizable character of the massless solution in a crucial way, can be simpler. Note that from the massive and massless GY equations
\begin{eqnarray}
&& \left\{-\frac{\partial ^2}{\partial r^2}-\frac{3}{r} \frac{\partial }{\partial r}+\frac{4l(l+1)}{r^2}+\mathcal{V}_{l,j}(r)+m^2\right\} \psi (r;m)=0, \label{appBmassiveeq}\\
&& \left\{-\frac{\partial ^2}{\partial r^2}-\frac{3}{r} \frac{\partial }{\partial r}+\frac{4l(l+1)}{r^2}+\mathcal{V}_{l,j}(r)\right\} \psi_0 (r)=0,
\end{eqnarray}
we can deduce the relation
\begin{eqnarray}
\frac{\partial }{\partial r}\left\{r^3 \psi '(r;m) \psi _0(r)- r^3\psi (r;m) \psi _0'(r)\right\}= m^2 r^3 \psi (r;m) \psi _0(r),
\end{eqnarray}
Integrating this equation from $r=0$ to $r=R_e$ and then dividing by $R_e^4 \psi_0'(R_e)$, we get
\begin{eqnarray}
\psi (R_e;m) \left(\frac{\psi '(R_e;m)}{\psi (R_e;m)} \frac{\psi _0(R_e)}{R_e \psi _0'(R_e)}-\frac{1}{R_e}\right)=m^2 \frac{\int _0^{R_e} r^3 \psi (r;m) \psi _0(r)dr}{R_e^4 \psi _0'(R_e)}. \label{bint}
\end{eqnarray}
For large $R_e$, we can write $\psi_0(R_e) \sim \frac{A}{R_e^3}$ (as $H(r)$ approaches 1 as $r\to\infty$) and $\psi(R;m) \sim B e^{mR_e}$, where $A$ and $B$ are some constants. Hence, by considering the $R_e\to\infty$ limit with (\ref{bint}), we are led to conclude that
\begin{eqnarray}
\lim_{R_e\to\infty} \psi(R_e;m) = -3m \lim_{R_e\to\infty} \frac{\int _0^{R_e}r^3 \psi (r;m) \psi _0(r)dr}{R_e^4 \psi _0'(R_e)}.
\end{eqnarray}
When mass $m$ is small, this equation simplifies to
\begin{eqnarray}
\lim_{R_e\to\infty} \psi(R_e;m) \sim -3m \frac{\int _0^\infty r^3 \psi _0(r)^2 dr}{\lim_{R_e\to\infty} [R_e^4 \psi _0'(R_e)]}. \label{bint2}
\end{eqnarray}
For our Case I the explicit massless solution is available from (\ref{Psi1sol}):
\begin{eqnarray}
\psi_{0,0}(r;m=0) = \psi_0(r) = \frac{1}{\left(r^{2\alpha}+1\right)^{\frac{3}{2\alpha}}}.
\end{eqnarray}
Using this solution, we can evaluate the expression appearing in the right hand side of (\ref{bint2}) explicitly. The result is finite, and in this way we can secure the following result for the related functional determinant:
\begin{eqnarray}
\frac{\det(\mathcal{H}_{0,0}+m^2)}{\det (-\partial_{(0)}^2 +m^2)} = \lim_{R_e\to\infty} \frac{\psi(R_e;m)}{\psi^{\text{free}}_0 (R_e;m)} \sim m \frac{\Gamma \!\left(1+\frac{1}{\alpha }\right) \Gamma \!\left(\frac{2}{\alpha }\right)}{2 \Gamma \!\left(\frac{3}{\alpha }\right)}. \label{appBmassivere}
\end{eqnarray}

We now turn to the case with $j=l\neq 0$, where the GY solutions are given by $2\times2$ matrices. First, note that we can approximate the potential in (\ref{largereq}) for small mass $m$ as
\begin{eqnarray}
\frac{4l(l+1)}{x^2}+\frac{1}{m^2} \mbox{\boldmath$\mathcal{V}$}_{l,l}\!\left(\frac{x}{m}\right) \longrightarrow \frac{1}{x^2} \left(
\begin{array}{cc}
 (2l+1)(2l+3) & 0 \\
 0 & (2l-1)(2l+1)
\end{array}
\right) +O\!\left(m^2\right).
\end{eqnarray}
Thus the leading order solution of (\ref{largereq}) using the matrix potential given above is
\begin{eqnarray}
\mbox{\boldmath$\varphi$}_0(x)=\left(
\begin{array}{cc}
 2 (2 l+2)! \left(\frac{2}{m}\right)^{2 l+1} \frac{I_{2 l+2}(x)}{x} & 0 \\
 0 & 2 (2 l)! \left(\frac{2}{m}\right)^{2 l-1} \frac{I_{2 l}(x)}{x}
\end{array}
\right) \mathbf{c}_0,
\end{eqnarray}
where $\mathbf{c}_0$ is a constant $2\times2$ matrix which can be determined using the condition (\ref{demand}) at $r=R$. To find $\mathbf{c}_0$, we need to know an asymptotic behavior of the massless GY solution. For this $j=l\neq 0$ the massless solution is given in (\ref{psilpre}), with relevant expressions also in (\ref{Psi1explicit}), (\ref{Psi2}) and (\ref{Phi3explicit}). Since this solution was found after a suitable unitary transformation was performed (with the help of the unitary matrix $U$ in (\ref{Udef})), our condition (\ref{demand}) translates to
\begin{eqnarray}
U^\dagger \mbox{\boldmath$\Psi$}_l(R) U \sim \mbox{\boldmath$\varphi$}_0(mR) \sim \left(
\begin{array}{cc}
 R^{2l+1} & 0 \\
 0 & R^{2l-1}
\end{array}
\right) \mathbf{c}_0.
\end{eqnarray}
As the asymptotic behavior of the determinant of $\mbox{\boldmath$\Psi$}_l(R)$ is given in (\ref{detpsi1asymp}), we find that
\begin{eqnarray}
\det \mathbf{c}_0= \frac{\det \mbox{\boldmath$\Psi$}_l(R)}{R^{4l}} = \frac{(2 l+1)^3 \Gamma \!\left(\frac{2 l+1}{\alpha }\right)^4}{8 l (l+1)^2 \Gamma \!\left(\frac{2 l}{\alpha }\right)^2 \Gamma \! \left(\frac{2 l+2}{\alpha }\right)^2}. \label{c0result}
\end{eqnarray}
The true asymptotic behavior of the determinant of the GY wave function can then be found as
\begin{eqnarray}
\lim_{r\to\infty} \det \mbox{\boldmath$\psi$}_{l,l}(r;m) \sim \lim_{x\to\infty} \det \mbox{\boldmath$\varphi$} _0(x)\sim \frac{2}{\pi } \frac{e^{2x}}{x^3} \left( \frac{2}{m}\right)^{4l} (2l+2)!(2l)! \det \mathbf{c}_0
\end{eqnarray}
with $\det \mathbf{c}_0$ given by (\ref{c0result}). Based on this, we obtain the following result:
\begin{eqnarray}
\frac{\det(\mbox{\boldmath$\mathcal{H}$}_{l,l} +m^2)}{\left\{\det(-\partial_{(l)}^2+m^2)\right\}^2} = \lim_{r\to\infty} \frac{\det \mbox{\boldmath$\psi$}_{l,l}(r;m)}{\psi_l^{\text{free}}(r;m)^2} &\sim& \frac{2l+2}{2l+1} \det \mathbf{c}_0. \label{psilsmallmlimit}
\end{eqnarray}
Here note that $\det \mathbf{c}_0$ corresponds to the result of the related massless functional determinant. Therefore (\ref{psilsmallmlimit}) apparently shows that the small mass limit of the functional determinant is different from the massless functional determinant. Only with the combination in (\ref{GammaL}), the small mass limit and exact zero mass analysis yield the same results.

\end{document}